\documentclass[aps,pra,superscriptaddress,twocolumn,notitlepage,10pt]{revtex4-1}
\usepackage{bm} \usepackage{graphicx} \usepackage{amsmath}
\usepackage{amsthm,stmaryrd,mathtools}
\usepackage{amssymb} \usepackage{epstopdf} \usepackage{txfonts}
\usepackage{xcolor}

\DeclareMathOperator*{\Tr}{Tr}

\DeclareMathOperator{\sign}{sign}

\DeclareRobustCommand\openzero{\leavevmode\hbox{0\kern-.55em0}}
\mathchardef\minus="002D


\newcommand{\ket}[1]{|{#1}\rangle}
\newcommand{\bra}[1]{\langle{#1}|}
\newcommand{\mean}[1]{\langle{#1}\rangle}

\newcommand\up\uparrow
\newcommand\down\downarrow

\begin{document}


\title{Gating classical information flow via equilibrium quantum phase transitions}

\author{Leonardo Banchi}
\affiliation{Department of Physics \& Astronomy, University College London (UCL), London WC1E 6BT, UK}

\author{Joaqu\'in Fern\'andez Rossier}
\affiliation{Quantalab,  International Iberian Nanotechnology Laboratory, 4715-330 Braga, Portugal}
\affiliation{Departamento de Física Aplicada, Universidad de Alicante, 03690 San Vicente del Raspeig, Spain}

\author{Cyrus F. Hirjibehedin}
\affiliation{Department of Physics \& Astronomy, University College London (UCL), London WC1E 6BT, UK}
\affiliation{London Centre for Nanotechnology, UCL, London WC1H 0AH, UK}
\affiliation{Department of Chemistry, UCL, London WC1H 0AJ, UK}

\author{Sougato Bose}
\affiliation{Department of Physics \& Astronomy, University College London (UCL), London WC1E 6BT, UK}

\date{\today}

\begin{abstract}

The development of communication channels at the ultimate size limit of atomic scale physical dimensions will make the use of quantum entities an imperative.
In this regime,  quantum fluctuations naturally become prominent, and are generally considered to be detrimental. Here we show that for spin-based information processing, these fluctuations can be uniquely exploited to gate the flow of classical binary information  across a magnetic chain in thermal equilibrium. Moreover, this information flow can be controlled with a modest external magnetic field that drives the system through different many-body quantum phases in which the orientation of the final spin does or does not reflect the orientation of the initial input. Our results are general for a wide class of anisotropic spin chains that act as magnetic cellular automata, and suggest that quantum phase transitions  play a unique role in driving classical information flow at the atomic scale.

\end{abstract}

\maketitle

As the size of information processing platforms decreases, quantum mechanics becomes more prominent, both as a resource \cite{nielsen2010quantum} and also as an intrinsic source of fluctuations \cite{devoret1995quantum,wong2002beyond}. 
%
%
%
Given the dramatic recent advances in the ability to  engineer finite structures of interacting atomic or molecular spins \cite{gambardella2002ferromagnetism,lee2004spin,hirjibehedin2006spin,kitchen2006atom,chen2008probing,khajetoorians2011realizing,khajetoorians2012atom,loth2012bistability,spinelli2014imaging,nadj2014observation,weber2012ohm,menzel2012information}, 
it becomes important to explore whether finite nano-scale chains of interacting quantum entities in thermal equilibrium are a viable on-chip connector for transmitting bits   using either their charge or spin degrees of freedom.  
However, despite the potential of these structures for 
low-dissipation spin-based information technology
\cite{khajetoorians2011realizing,menzel2012information,loth2012bistability}, 
there has been no investigation of the capacity of their robust thermal (static) states to
 convey classical information, in particular, in the sense of Shannon's quantitative 
theory \cite{shannon1948amathematical}. 
Moreover, in the atomic regime, quantum fluctuations, rather than being merely  noise, 
play a fundamental role and 
can actively drive a phase transition in a many-body system 
\cite{dutta2015quantum,sachdev2007quantum}. 
Here we
show that these quantum phase transitions can produce striking 
changes in the information transfer capacity and  thereby
demonstrate a fully quantum methodology for gating the classical information 
flow through a large class of magnetic chains.

\begin{figure}[h]
\centering
\includegraphics[width=0.45\textwidth]{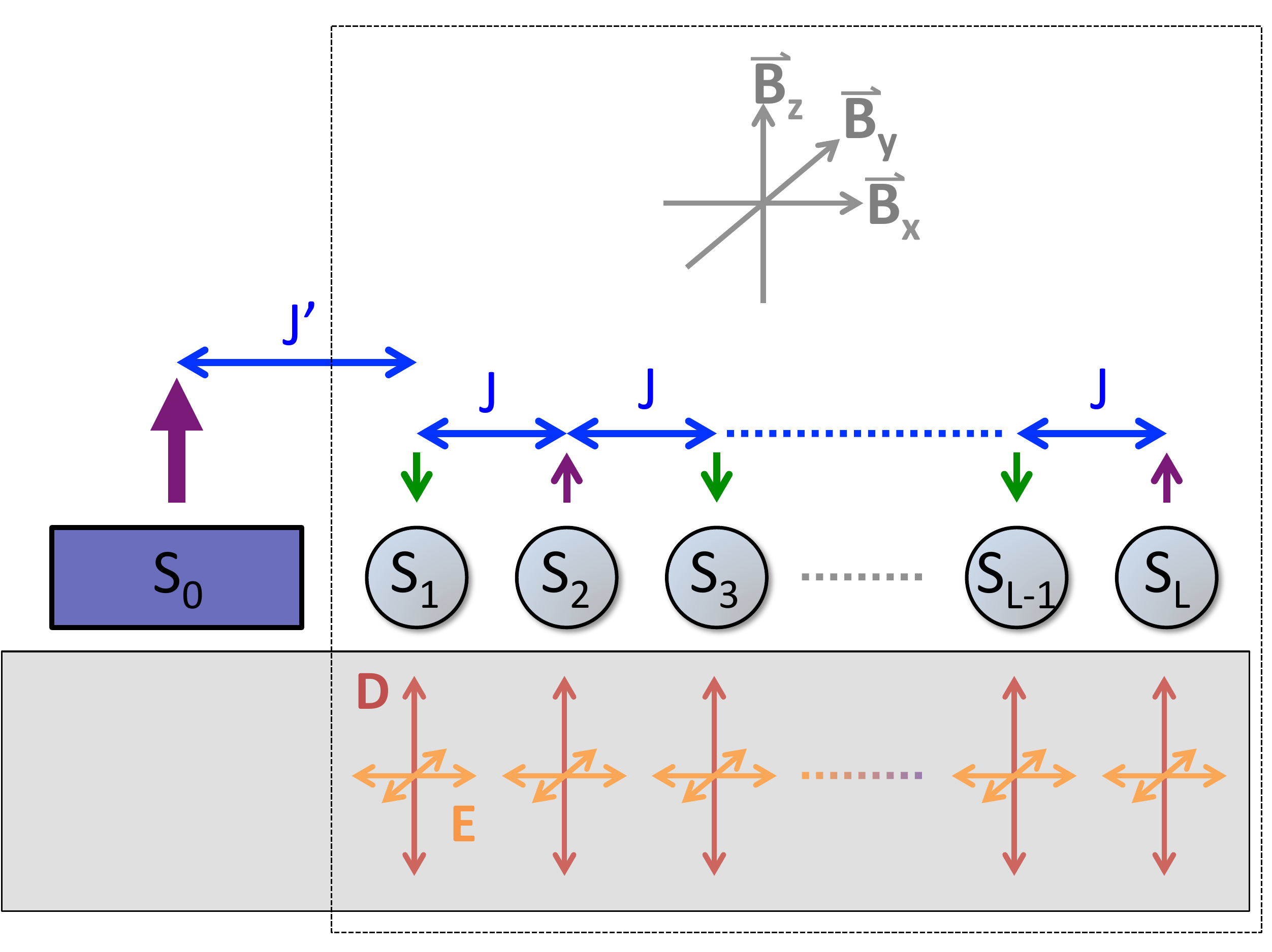}
\caption{ { A spin chain and its interactions on a surface.}
Schematic illustrating individual atomic spins $S_1$ ... $S_L$ (gray balls) in a chain on a surface (solid grey rectangle) coupled by an interaction with strength $J$ (blue arrows). The first spin $S_1$ is coupled to a large, semi-classical spin $S_0$ (purple rectangle) via an interaction $J'$ (blue arrow). Also shown are the axial and transverse anisotropy terms, $D$ (red arrows) and $E$ (orange arrows) respectively, and the components of an externally applied magnetic field $\vec B$ (gray arrows). Green and purple arrows indicate the magnetic orientation of the spins for an antiferromagnetic coupling. 
  }
  \label{fig:cartoon}
\end{figure}

We consider a generic setup for transmitting classical digital information through the equilibrium state of a quantum spin chain (Fig.~\ref{fig:cartoon}). As in recent experiments \cite{khajetoorians2011realizing}, the magnetic island on the left has uniaxial anisotropy so that it only has two ground states and therefore encodes one bit of classical information. This input is inserted into a spin chain via exchange coupling to the first quantum spin $S_1$. Every other  spin $S_j$ of the chain is coupled to its nearest neighbors. The magnetic island is  sufficiently large as to be described with a classical magnetization $S_0$, 
and to make the back action of the quantum spins negligible. The logical state  of the island can be controlled  independently, for example by using external magnetic pulses \cite{khajetoorians2011realizing}. In this system the output is defined by the orientation of the last spin $S_L$, where $L$ counts the number of quantum spins in the  the chain.  The readout can be realised using spin-polarised tunnelling \cite{wiesendanger2009SPSTMreview}.
Typically the initialization and measurement times are much slower than the equilibration time, so during the output measurement the chain is in its equilibrium state. 

If we ignore both quantum and thermal fluctuations, so that the spin chain is described with classical Ising  spins at $T=0$, with two equivalent ground states, perfect transmission occurs from the island to the opposite boundary: fixing the logical state of the first spin $S_1$ selects one of the two ground states for the entire chain. It can be readily seen that thermal fluctuations, at the classical level, destroy this ideal picture.  To quantify this, we make use of Shannon's seminal work \cite{shannon1948amathematical}, where he showed that the maximum rate at which information can be transmitted over a memory-less communications channel with arbitrarily small decoding error is given by the so-called channel capacity $C$ (Supplementary Information). The capacity is 1 for error-free channels, 0 for fully broken ones, and in some simple cases is a function of the  probability $P_f$ of bit-flip errors at the output. Applying this formalism to the classical Ising model at inverse temperature $\beta=1/(k_B T)$ (Supplementary Information), where each spin can point either up or down, we find $P_f=0$ at $T=0$ and the chain perfectly transfers information. However, at high temperature the capacity goes down as $C\simeq \exp(-4 L e^{-2\beta J})$, where $J$ is the strength of the Ising coupling. Therefore, thermal fluctuations limit the maximum length of the chain for reliable information transfer to $L\approx e^{2\beta J}$.

Whereas it is possible in principle to suppress classical fluctuations by reducing $T$, the same will not be true for quantum spin fluctuations, making a study of their effects imperative. Once we switch from classical to quantum systems, the reconstruction of the output density matrix is necessary to determine the ultimate rate of classical information transfer \cite{holevo1973bounds,giovannetti2005information,giovannetti2014ultimate,giovannetti2013electromagnetic,banaszek2012quantum} (see also the Supplementary Information). However, motivated by an experimental perspective in adatom chains, we focus on a simpler scheme where the {\it digital} output is encoded into the sign of the magnetization $\mean{S_L^z} $ of the last spin, as described above, and we study
the information capacity when the channel spins are described with the anisotropic Heisenberg  Hamiltonian:

\begin{align}
  \mathcal H_{\rm chain} &= J \sum_{n=1}^{L-1} \vec S_n\cdot \vec S_{n+1} + 
  \sum_{n=1}^L \mathcal H_n ~, 
  \label{e.Ham}\\
  \mathcal H_n &= D(S_n^z)^2+E[(S_n^x)^2-(S_n^y)^2] + \vec B \cdot \vec S_n~,
  \label{e.anis}
\end{align}
where (Fig.~\ref{fig:cartoon}) $L$ is the length of the chain, $S^\alpha_n$ is the quantum spin operator along the direction $\alpha={x,y,z}$ acting on the $n$-th spin, $\vec B$ is the magnetic field (with the Bohr magneton and the Land\'e g-factor absorbed into the definition of $\vec B$), $D$ is the (axial) zero field splitting, $E$ is the planar (transverse) anisotropy of the crystal field interaction, and $J$ is the exchange integral between neighboring sites.  This Hamiltonian can describe a variety of different physical phenomena, depending on the relative value of the exchange interaction and anisotropy, as well as the value of the spin $S$. Importantly, these quantities can be tuned experimentally and the model has been used to successfully describe experimentally realised spin chains \cite{hirjibehedin2006spin,chen2008probing,loth2012bistability,spinelli2014imaging}.

The effect of a large magnetic island, whose orientation can be tuned externally \cite{khajetoorians2011realizing}, can be described by a classical magnetic field 
$B_0=J'S_0$ pointing along the $z$-direction, so that the total Hamiltonian is $  \mathcal H = \mathcal H_{\rm chain} + B_0 S^z_1$. Here we will focus on antiferromagnetic (AFM) chains (i.e. $J>0$), which have shown considerable stability at low temperatures \cite{loth2012bistability}. Although we find that FM systems generally have a higher capacity than their AFM counterparts, this capacity is highly sensitive to external fields in the $z$-direction (Supplementary Information).

\begin{figure}[th]
  \centering
  \includegraphics[width=0.45\textwidth]{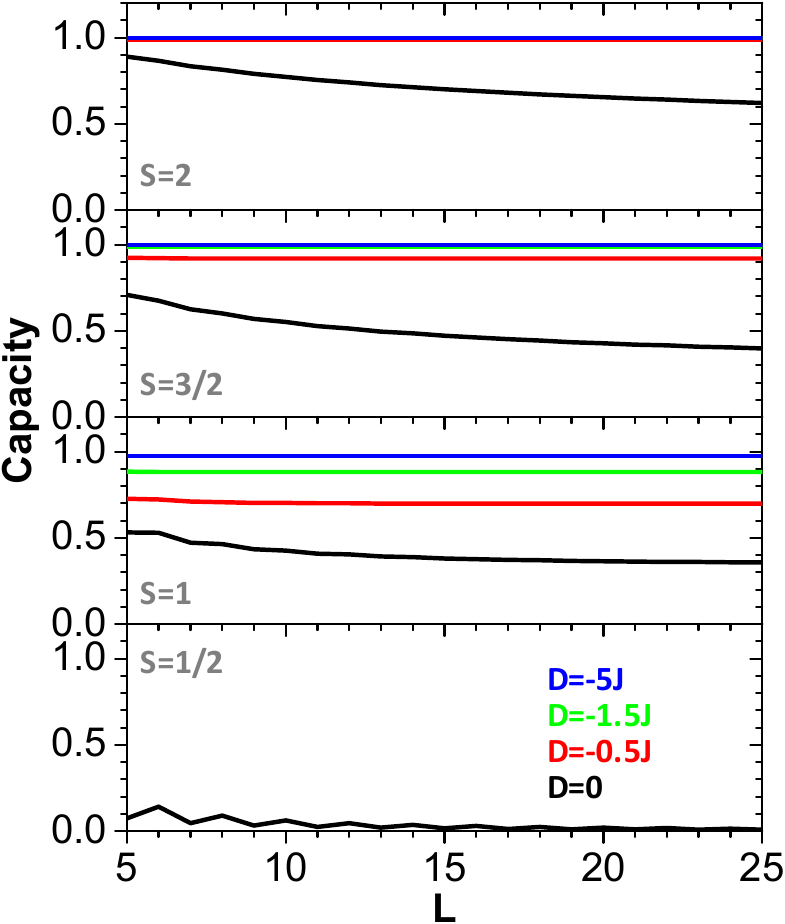}
  \caption{ { Impact of axial anisotropy on channel capacity.}
  Capacity vs. chain length for Heisenberg-coupled spin chains with $E=0$ and $B_0=100 J$ for $S=2$, $S=3/2$, $S=1$, and  $S=1/2$, as calculated using DMRG for $D=0$ (black), $D=-0.5J$ (red), $D=-1.5J$, and $D=-5J$ (blue). 
  }
  \label{fig:length}
\end{figure}

We first consider a chain with isotropic spin-spin couplings, with $\vec B=E=D=0$. Figure \ref{fig:length} shows the digital channel capacity (Supplementary Information) of such finite length quantum spin chains, where we use a DMRG algorithm \cite{schollwock2011density} to compute the ground state.  It is apparent that the channel capacity is smaller for chains of lower spins, with isotropic $S=1/2$ systems being particularly poorly suited for information transfer. Even for larger spins, the capacity clearly decreases as the length of the chain increases (Fig. \ref{fig:length}) because the perturbation due to the local coupling with the island, which breaks the rotational symmetry of the state of the spin chain, is not able to propagate along the chain. 

To stabilise only the two states $\ket{m^z_n{=}{\pm} S}$, it is therefore natural to consider spin systems with $S\ge1$ and large negative $D$. As seen in Fig. \ref{fig:length}, as we increase the uniaxial  anisotropy,  the channel capacity increases and becomes less dependent on the channel length. For sufficiently large uniaxial anisotropy  ($D = -5 J$, $E=0$) the channel is similar to  classical Ising spins at $T=0$. Therefore, both temperature and antiferromagnetic flip-flop interactions driven by exchange can compromise the channel capacity.

\begin{figure}[th]
  \centering
  \includegraphics[width=0.45\textwidth]{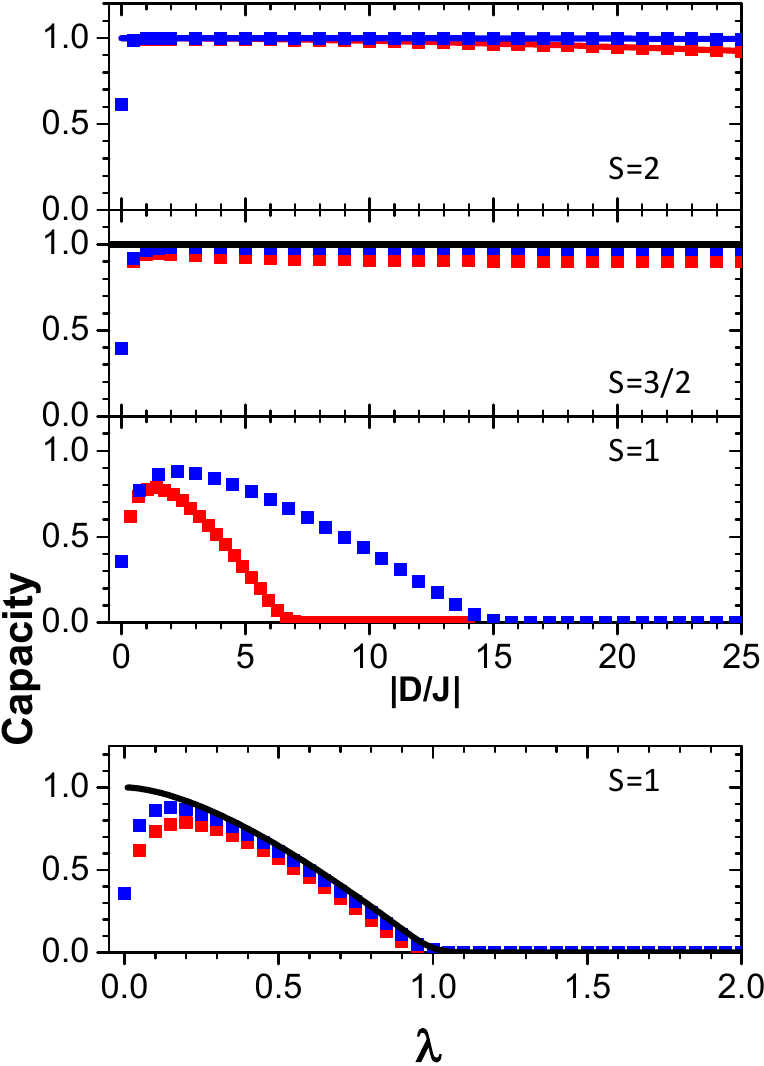}
  \caption{ { Impact of quantum fluctuations on spin chain capacities.} 
    (upper panel) Capacity of a spin chain with $L=25$, $B_0=J$, and $\vec{B}=0$, as obtained from DMRG calculations for $S=2$, $S=3/2$ and $S=1$, plotted as a function of $|D/J|$ for $D=-7E$ (red squares) and $D=-15E$ (blue squares). The ratio $\eta=D/E$ is kept fixed as $|D/J|$ increases. 
    (lower panel) The above results for $S=1$ are plotted against $\lambda$. 
    Solid lines, in both panels, show the predictions of the 
    effective model (Table~\ref{tab:hameff}) with 
    the same parameters. A single black line is used when the results are independent of 
    $\eta$. 
  }   
  \label{fig:fluctuations}
\end{figure}

We now show that there is yet another factor that affects  the channel capacity decisively: quantum spin tunnelling of the individual spins, driven in this system by 
the transverse anisotropy \cite{delgado2012storage} $E$. As shown in Fig. \ref{fig:fluctuations}, 
where we keep $E/D$ fixed and vary $D/J$, the capacity remains relatively large (above 0.9) for $S=3/2$ and $S=2$, even with finite values of $E$ once a sufficiently large ($|D| \gtrsim J$) axial anisotropy is present. In sharp contrast, however, the case of $S=1$ (Fig. \ref{fig:fluctuations}) exhibits a rapid decrease in capacity down to zero above a critical value.
The  role played by the in-plane anisotropy $E$, very different from the effect of the uniaxial anisotropy $D$,  can be understood analytically, in the limit $|D/J|\gg 1$ and $D<0$. In this case,  each spin $S\ge1$ is approximated \cite{delgado2014emergence} with an effective two-level system  $\ket{{{\pm}} S}$, namely with a pseudo-spin $\tau=1/2$. At the single spin level, Kramers theorem ensures that these two levels are degenerate for half-integers spins, but are in general split due to single-spin quantum spin tunnelling (SS-QST) in the case of integer spins. This splitting acts as an effective field in the pseudospin space, so that the resulting model for the channel is the quantum Ising model (QIM) in a transverse field $\lambda$ (see Table~\ref{tab:hameff}), which is one of the paradigmatic systems for the study of quantum phase transitions \cite{dutta2015quantum}. The QIM has two distinct phases. The interaction dominated phase, with $|\lambda|<1$, has two ordered ground states, characterised in the thermodynamic limit ($L\to\infty$) by long-range correlations, and a non-zero order parameter $\langle \tau^z_n\rangle$. For $|\lambda|>1$, the field dominated phase, there is a unique paramagnetic ground state that results in a quantum disordered phase with short-range correlations and a vanishing order parameter.


\begin{table}
  \centering
  \begin{tabular}{|c|c|c|}
    \hline
    \multicolumn{3}{|c|}{ 
      $ \mathcal H_{\rm eff} = \mu \tau_1^z + \sum_n\left(\lambda \tau_n^x +
      \tau_n^z\tau_{n+1}^z\right) $
    }
    \\
    \hline 
    \hline
  ~ ~ Spin 1 ~ ~ & ~ ~ ~ ~ ~ $\mu\simeq\frac{B_0}{J}$ ~ ~ ~ ~ ~ & 
  ~ ~ ~ ~ ~ $\lambda \simeq \frac{E}{J}-\frac{B_x^2}{2|D|J}$ ~ ~ ~ ~ ~  
  \\
    \hline
  {Spin }$\frac32$ & $\mu\simeq\frac{2B_0}{3J}$ & $\lambda \simeq
  -\frac{2B_xE}{3|D|J}+
  \frac{B_x^3}{12 D^2 J}+\frac{B_x E^2}{3 D^2 J} 
  $
  \\
    \hline
  {Spin }2& $\mu\simeq\frac{B_0}{2J}$ & $\lambda \simeq -\frac{3E^2}{8|D|J}+\frac{5 B_x^2 E}{24 D^2 J}-\frac{B_x^4}{96 |D|^3 J}$
  \\
    \hline
  \end{tabular}
  \caption{  
  Low-energy effective Hamiltonian for anti-ferromagnetic chains in the subspace $\ket{m_n^z{=}{{\pm}}S}$ 
  when $D<0$, ${\vec B} \text{ } {\parallel} \text{ } {\hat x}$ and 
  $\vec B_0 \text{ } {\parallel} \text{ } \hat z$. 
  The operators $\tau^\alpha_n$ are effective
    spin-$1/2$ Pauli operators acting on the low-energy subspace. 
    The effective theory has been obtained with the theory presented in
    Ref \cite{jia2015integrated} using a second order expansion in $\epsilon=(D/J)^{-1}$ (third order for $S=2$), assuming that 
    $(E/J) \simeq \mathcal O(\epsilon^\eta)$, 
    $(B_x/J) \simeq \mathcal O(\epsilon^\beta)$ where $0<\eta<1, 0<\beta<1$ and keeping only 
    the dominant terms. Within these assumptions, no $\epsilon^2$ corrections are present for $S=1$.
  }
  \label{tab:hameff}
\end{table}

Within the QIM, we can find an analytical expression (Supplementary Information) for the  magnetization of the output spin after the local perturbation introduced by the island: 
\begin{align}
  m^z_L \equiv \mean{\tau^z_L} \simeq \begin{cases}
    {\rm sign}(\mu)(-1)^{L{+}1}\sqrt{1-\lambda^2}~ & {\rm if ~} |\lambda|<1~,
    \\
    0 & {\rm if ~} |\lambda|>1~, \\
  \end{cases} 
  \label{e.magn}
\end{align}
when $L\gg1$, where $\mu\propto \frac{B_0}{J}$ (see Table \ref{tab:hameff}). Equation \eqref{e.magn} shows that in the thermodynamic limit $m^z_L$ is a non-analytic function of $\mu$ and  depends only on its sign, while for finite systems this non-analytic behaviour is smoothened. 
Finite-size effects are negligible far from the critical point, and can be estimated from conformal invariance at $\lambda\approx1$ \cite{igloi1997density,burkhardt1991density}.
From Eq. \eqref{e.magn}, the capacity $C$ can be evaluated and we find in the thermodynamic limit that $C=1+\sum_{\pm} \left(\frac{1\pm\sqrt{1-\lambda^2}}2\right) \log_2\left(\frac{1\pm\sqrt{1-\lambda^2}}2\right) $ for $|\lambda|<1$, while the capacity is zero when $m_L^z=0$. This shows that the chain acts as a ``wire'' able to carry digital information from the island to the distant opposite end only in the ordered phase, while when $|\lambda|>1$ the appearance of a unique gapped ground state blocks the information flow. 

The dependence of $\lambda$ on the physical parameters of the chain is very different for $S=1$ 
(Table \ref{tab:hameff}), and can be used to account for the very different behaviour shown in Fig. \ref{fig:fluctuations}. In particular, for $S=1$ as $\lambda\simeq \frac{E}{J}$ is increased, the system undergoes a quantum phase transition. 
In contrast, for $S=3/2$ there is no zero field splitting, in agreement with Kramers theorem, and $\lambda=0$ for $\vec B=0$. For $S=2$, $\lambda\simeq\frac{E^2}{DJ}$, which is much smaller than for $S=1$.  As a result, for fixed $E/D$, the fluctuation dominated paramagnetic phase can only be achieved when $D/J$ is very large, $D/J>(E/D)^{-2}$ -- see also Supplementary Section II.B and Fig.~S1. As seen in Fig. \ref{fig:fluctuations}, the effective model presented in Table \ref{tab:hameff} is in excellent agreement with the DMRG results (for $|D| \gtrsim J$), showing that $S=1$ systems are unique in providing a flexible system where the flow of information depends on quantum phases. 




\begin{figure}[t]
  \centering
  \includegraphics[width=.45\textwidth]{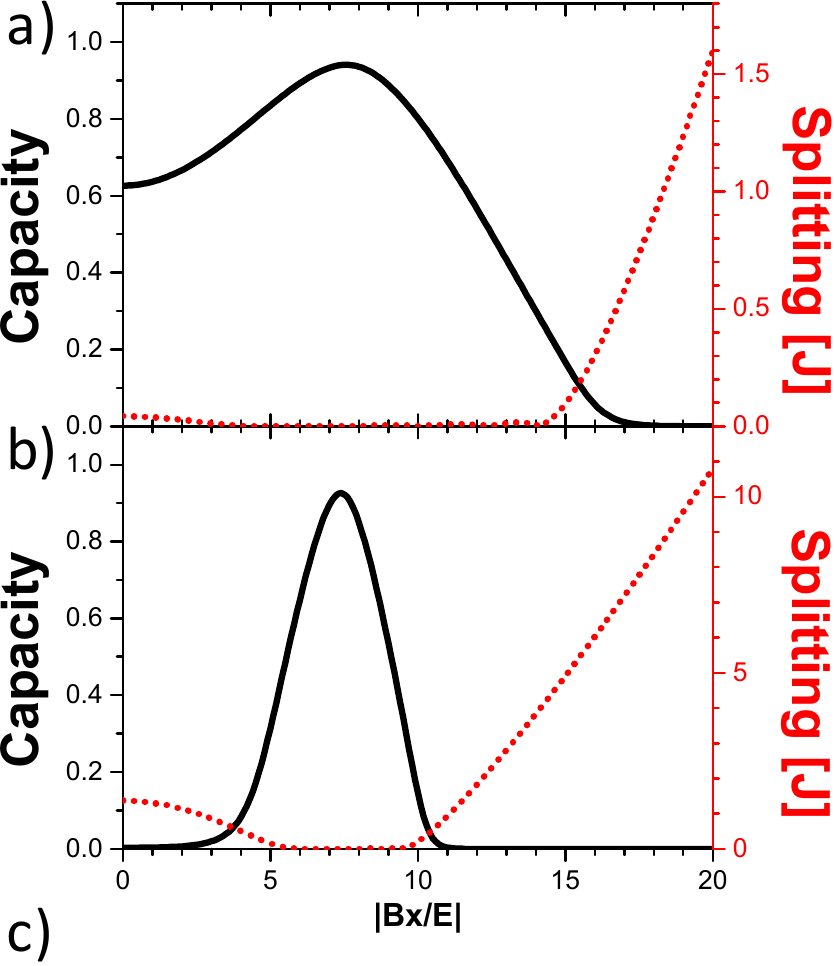}
  \includegraphics[width=.45\textwidth]{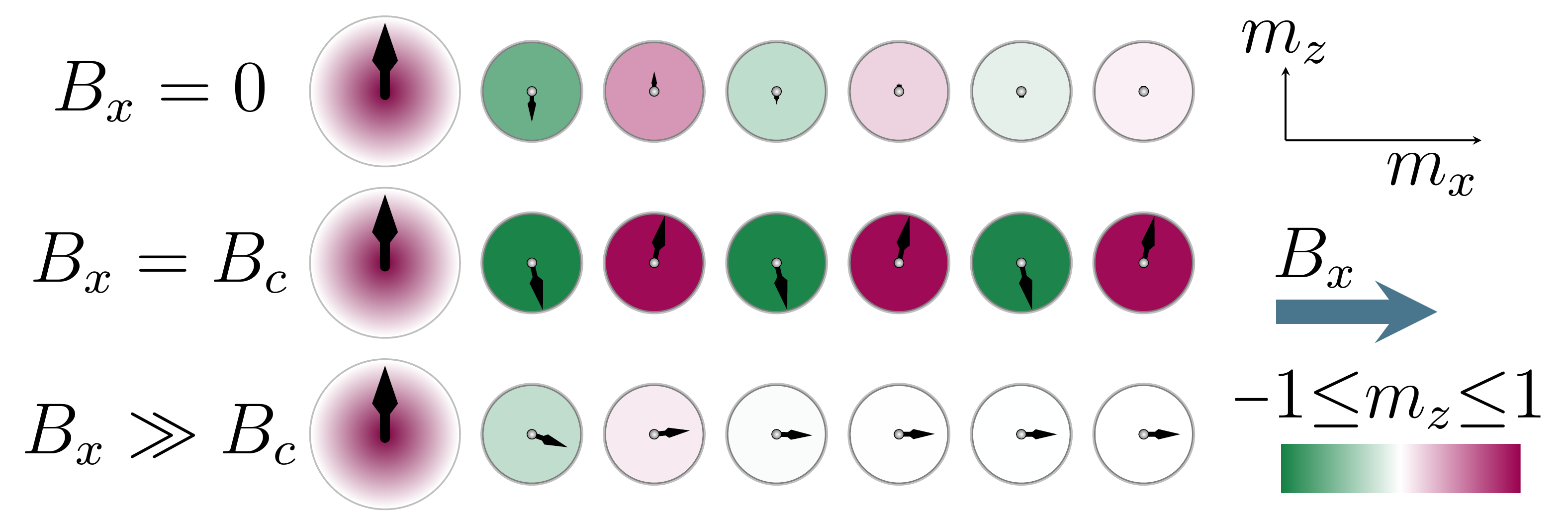}
  \caption{{Gating capacity via quantum fluctuations with an externally applied magnetic field.} 
  a) Capacity (solid black line) and splitting between the two lowest energy states (dotted red line) for an $S=1$ spin chain with with $L=6$, $D=-25E$, and $E=0.5J$, as obtained from exact diagonalisation, plotted as a function of $|B_x/E|$. The calculation for the capacity uses $B_0=J$, while the calculation for the gap uses $B_0=0$. b) Same as panel (a) with $E=1.5J$, so that quantum fluctuations completely destroy the capacity at $B_x=0$. c) Spin configurations in the $xz$-plane for a chain with the same parameters as for panel b. The initial, larger spin represents the magnetic island.}
  \label{fig:field}
\end{figure}

Having unveiled the relevant role played by quantum fluctuations and SS-QST 
permits us to devise a strategy to control the spin chain channel capacity using externally tunable parameters. For this, we use the fact that, 
at the single-spin level, the
application of a magnetic field $B_x$ along the hard-axis 
modulates the quantum spin tunnelling splitting \cite{garg1993topologically,wernsdorfer1999quantum}, which acts as an effective magnetic field $\lambda$ in the pseudospin space, and thereby also affects the collective state in the spin chain. This is seen in Fig. \ref{fig:field} where we consider a spin chain with $S=1$ at two different values of the anisotropy: $E=0.5J$ in Fig.~\ref{fig:field}a, and $E=1.5J$ in Figs.~\ref{fig:field}b,c.  Without applied fields, the chain of Fig.~\ref{fig:field}a is in the ordered phase ($\lambda=0.5$) with an imperfect capacity, and that of Fig. \ref{fig:field}b is in the quantum disordered phase ($\lambda=1.5$) with vanishing channel capacity. 
In this non-trivial regime, as seen in Fig. \ref{fig:field}c, the AFM configuration is preserved only around the magnetic island while far from this boundary all the spin components $m^\alpha_n$ are zero, meaning that each spin is highly entangled with the others \cite{amico2008entanglement}. 
Moreover, at $\lambda=1$ the magnetization profile along the chain 
can also be obtained from
conformal invariance \cite{igloi1997density,burkhardt1991density}. 
As we start increasing $B_x$, the SS-QST decreases in both cases and the channel capacity is improved. This effect is even more dramatic in Fig. \ref{fig:field}b) where, in an initially disordered chain, $\lambda(B_x)$ is decreased below the critical value $\lambda=1$ and magnetic order along the $z$ axis is established due to application of a magnetic field along the orthogonal $x$ axis. At the optimal field $B_c$, it is $\lambda=0$ and
the channel capacity is maximal (see also Fig.~\ref{fig:field}c). 
As the magnetic field $B_x$ is further increased, the system eventually reaches the trivial limit where the Zeeman energy dominates all other energy scales, and
the capacity to transmit information vanishes.
 Our numerical calculations show that the dramatic effect of the transverse  magnetic field on the channel capacity is remarkable for anisotropic $S=1$ chains, the only ones for which $\lambda$ is independent of the uniaxial anisotropy $D$ at $B_x=0$.  The channel capacity of chains with $S=3/2$ spins shows a weak dependence on $B_x$, expected from the functional dependence of $\lambda(B_x)$ (see Table \ref{tab:hameff}), 
while in higher-spin systems only the regime $\lambda\simeq 0$ is accessible due to the damping of the QST splitting for large $|D|$. 
To the leading orders in perturbation theory (see Supplementary Information for more general arguments), 
maximal capacity ($\lambda=0$) corresponds to the ``diabolic points'' \cite{wernsdorfer1999quantum} where the QST splitting of each spin vanishes. On the other hand, in a chain the energy splitting is small (vanishes in the thermodynamic limit) in the whole interval $|\lambda|<1$ (see Fig.~\ref{fig:field}a,b), 
so the {\it regions} of non-zero capacity correspond to quantum ordered chains with (almost) degenerate ground states.

The theoretical framework introduced here uses the standard metric from
information theory for the first time to analyze the transfer of classical
information along a quantum magnetic chain in thermal equilibrium. 
Our approach highlights the deep relationship between quantum magnetic phases and information transmission. 
Thus we find a way to take advantage of the competition between various energy scales -- such as uniaxial and in-plane anisotropy as well as exchange interactions and Zeeman coupling -- to devise a non-trivial method to tune magnetic order, and thereby the channel capacity in the spin chain, 
using external magnetic fields.
%
Given the typical values of anisotropy and spin coupling that have been observed for magnetic atoms on surfaces \cite{gambardella2002ferromagnetism,gambardella2004giant,hirjibehedin2006spin,hirjibehedin2007anisotropy,khajetoorians2011realizing,khajetoorians2013spin,bryant2013local,spinelli2014imaging} ($|D| \sim 0.1 - 1 \text{ meV}$, $|E| \leq |D|/3$, and $|J| \sim 0.1 - 10 \text{ meV}$), $B_c$ can be below 10 T and therefore easily accessible with current experimental probes. 
Future work will consider the additional effects of electronic coupling of the spins in the chain to the substrate, which can be tuned by fabricating the structures on top of superconducting \cite{heinrich2013protection} or even insulating materials, the latter of which would require the use of force-based microscopy for the fabrication and readout. Our results
also suggest that more complex quasi-one-dimensional spin structures, such as
spin chains with next nearest neighbor coupling or spin ladders, may also exhibit intriguing phenomena that can be used to manipulate their classical information capacity.
Finally, we note that STM-based pump-probe techniques \cite{loth2010measurement} may be extended in the future with multiple tips to enable experimental exploration of the propagation of the dynamics along the chain.


\begin{acknowledgements}
  {\it Acknowledgements}:--
We thank Ian Affleck, Cristian Batista, Adrian Feiguin, Andrew Fisher, Chris Lutz, and Sebastian Loth for stimulating discussions, some of which were during SPICE workshop ``Magnetic adatoms as building blocks for quantum magnetism" and the MPI-IBM workshop in Almaden.
L.B. and S.B. acknowledge financial support from the ERC (Starting Grant 308253 PACOMANEDIA); C.F.H. from the Leverhulme Trust (RPG-2012-754) and EPSRC (EP/M009564/1); J.F.R. from the UCLQ Visitors Programme.
\end{acknowledgements}

\renewcommand{\thetable}{S\arabic{table}}%
\renewcommand{\thefigure}{S\arabic{figure}}
\renewcommand{\theequation}{S\arabic{equation}}

\clearpage
\appendix

\section{Channel capacity}
\label{a:capacity}
To formally define the capacity of a classical channel, the input set is $X=\{0,1\}$ where $0$ and $1$ refer to positive and negative magnetization respectively. Next, we call $Y=\{0,1,\rm{E}\}$ the output set where $0$ and $1$ refers to positive and negative magnetization $\mean{S_L^z}$ while $E$ is an error where the sign of the magnetization cannot be specified, e.g. when the magnetization is zero. Assuming that the channel is memoryless (i.e. independent of the prior use), for the sake of the error analysis all input messages can be specified with a probability distribution $p_{\rm in}(x)$ over the inputs $x\in X$, e.g. the concentration of bits $0$ in the message. The resulting output distribution  $p_{\rm out}(y)$, for $y\in Y$ depends on the input one and on  the conditional probability $p(y|x)$ to obtain the outcome $y$ given that the input is $x$. The mutual information between the input and the output distributions is given by 
$ 
  I(X:Y) = H(Y) - H(Y|X),
  $
where $H(Y)=-\sum_y p_{\rm out}(y) \log_2 p_{\rm out}(y)$ is the Shannon entropy and $H(Y|X) = \sum_x p_{\rm in}(x) H(Y|X{=}x) = -\sum_x p_{\rm in}(x) \sum_y p(y|x) \log_2 p(y|x)$. The channel capacity is then defined by \cite{mackay2003information}
\begin{align}
  C=\max_{p_{\rm in}} I(X:Y)~. 
  \label{e.capacity}
\end{align}
For a binary input $p_0=p$ and $p_1=1-p$, so the above equation can be solved numerically by maximizing over the single variable $p$. Therefore, the only quantities that we have to obtain from the model are the conditional probabilities $p(y|x)$. 

In our model, $x$ refers to the orientation of the magnetic island while $y$ is the sign of the last spin in the chain along the $z$ direction. Therefore, setting the applied field to either $\up$ or $\down$, one can get the conditional outcome
\begin{align}
  p(0|x) &= \sum_{m^z>0} \bra{m^z}\rho_L\ket{m^z}~,\\
  p(1|x) &= \sum_{m^z<0} \bra{m^z}\rho_L\ket{m^z}~,\\
  p(\rm{E}|x) &= 1-p(0|x) -p(1|x) ~,
  \label{e.condprob}
\end{align}
where $\ket{m^z}$ are the eigenstates of $S^z$, i.e. $S^z=\sum m^z\ket{m^z}\bra{m^z}$ and $\rho_L$ is the reduced density matrix of the last spin of the chain. Clearly $p(\rm{E}|x)$ is different from zero only for integer spins. However, this definition can be extended also to consider experimental imperfections in detecting the sign of the last spin. 

When we can neglect the error outcome (namely $p({\rm E}|x)=0$) and there is the symmetry $p(1|0)=p(0|1)=P_f$, $p(0|0)=p(1|1)=1-P_f$, then the channel implements a {\it binary symmetric channel} whose capacity can be evaluated analytically \cite{mackay2003information} as $H[P_f]$, where $P_f$ is the probability of bit-flip errors and $H[p] = -p\log_2p-(1-p)\log_2(1-p)$ is the binary entropy function. 
However, when the channel is not symmetric its capacity cannot be specified with a single parameter $P_f$, because there is different bit-flip probability depending on whether the input was either 0 or 1. For instance, in a fully polarized FM chain in the state $\ket{\uparrow\uparrow\uparrow\dots}$ the states $\ket\uparrow$ are perfectly transferred while the states $\ket\downarrow$ are never transferred. This simple argument shows that the channel capacity is the most natural quantity to look at in the general case.

\section{Solution of the effective model}
\subsection{ Analytic solution}
\label{a:fermi}
The Ising model with both longitudinal and transverse  magnetic fields is a paradigmatic model for non-integrable systems \cite{kim2013ballistic}. However, since in our case the longitudinal field is only on the first site, we can map the chain of Table \ref{tab:hameff} to a quadratic fermionic Hamiltonian with a linear perturbation. Following Kitaev \cite{kitaev2001unpaired,kitaev2009topological}, the Hamiltonian of Table \ref{tab:hameff} can be cast into a chain of Majorana fermions via the Jordan-Wigner transformation, $w_n= \prod_{j<n}(-\tau_j^x)\tau_n^y$, $w_{L{+}n}= \prod_{j<n}(-\tau_j^x)\tau_n^z$, and $w_nw_{L{+}n}= i\tau_n^x$. The $2L$ fermionic real operators $w_j$ satisfy the algebra $\{w_j,w_k\}=2\delta_{jk}$ and the Hamiltonian $\mathcal H_{\rm eff}$ takes the form
\begin{align}
  \mathcal H_{\rm eff} = \bar \mu^T\bar w + \frac{\bar w^T\Lambda\bar w}2~,
  \label{e.Hmajor}
\end{align}
where the bars denote vectors with $2L$ components, $\mu_j = \mu\delta_{j,L{+}1}$, $\Lambda=\begin{pmatrix}  0 & A^\dagger \\ A & 0 \end{pmatrix}$, where $A_{jk} = i \lambda \delta_{jk} + i \delta_{k,j{+}1}$. Unlike the Kitaev Hamiltonian \cite{kitaev2001unpaired}, Eq.\eqref{e.Hmajor}, has also linear term which cannot be removed with a Bogoliubov transformation; such transformations can remove the linear term only when $\mu$ is a Grassmann variable \cite{blaizot1986quantum}, an unphysical case in spin systems. In order to diagonalize Eq.\eqref{e.Hmajor} we define the unitary transformation $\mathcal U=e^{\sum_k i g_k w_k/2}$ and we want to find the vector $g_k$ which removes the linear term from the Hamiltonian, namely 
$
  \tilde{\mathcal H}_{\rm eff} = \mathcal U \mathcal H_{\rm eff} 
  \mathcal U^\dagger  = 
  \sum_{ij} \tilde\Lambda_{ij} w_i w_j/2~.
$
After  long but simple passages, we find that the above requirement is satisfied if $g$ satisfies the linear equation 
$  \Lambda g = -i\frac{\|g\|}{\tan\|g\|}\mu~.
$ 
This system of equations is solved by evaluating  the normalized solution of $\Lambda\tilde g = -i\mu$, which we call $\tilde g$,  and then finding the normalization factor such that $g = \|g\| \tilde g$. We find that
$  \tan \|\bar g\| = \frac{\|\bar \mu\|^2}{|\bar \mu^T\Lambda\tilde g|}~,
$ 
and that the renormalized Hamiltonian matrix $\tilde\Lambda$ is 
\begin{align}
  \tilde\Lambda_{ij} = \Lambda_{ij} +  i\tan\|g\| \frac{\tilde g_i \mu_j-\mu_i\tilde g_j}{1+\sqrt{1+\tan^2\|g\|}}~.
  \label{e.Gfinal}
\end{align}
Since $\tilde{\mathcal H}_{\rm eff}$ is quadratic without linear terms it can be diagonalized using standard methods \cite{kitaev2001unpaired}. In particular, $\tilde \Lambda$ can be cast into the canonical form via an real orthogonal transformation \cite{kitaev2001unpaired} $W$, namely $W^T\tilde\Lambda W=\Omega\otimes\sigma^y$ being $\Omega$ a real diagonal matrix whose diagonal elements $\omega_n>0$ are the eigenfrequencies of the Hamiltonian. 

The expectation value $\tau_L^z$ is obtained from the relation
$\tau_L^z=iPw_L$, where
$P=\prod_j(-\tau_j^x)=i^L\prod_{n=1}^{2L} w_n$ 
is the parity of the chain. We find 
$
m^z_L = \langle\tau_L^z\rangle = \tilde g_{L} p \sin\|\bar g\|~,
$
where $p$ is the parity of the transformed chain, which is the only quantity that depends on the initial state. When the chain is in its ground state $p={\rm sign}(\det W)=\pm1$, while in the thermal case $|p| = \prod_{k=1}^L\tanh[\beta\omega_k] <1$.

Our analytical treatment unveils many important informations. The  vector $\tilde g$ is independent of strength $\mu$ of the local field, being normalized, so the dependence upon the strength is encoded only in $\|g\|$. To make this dependence more explicit we define $\phi$ such that $\Xi=(\tan\|\bar g\|)/\|\bar \mu\|=(\tan\|\bar g\|)/|\mu|$ where 
$\tan\phi$ does not depend on $\mu$. Therefore, 
\begin{align}
  m^z_L \propto \frac{\mu\Xi}{\sqrt{1+\mu^2\Xi^2}}~. 
  \label{e.mzbeta}
\end{align}
The above equation shows the dependence of the magnetization of the $L$th spin on the applied field on the first spin. It holds beyond linear response theory, being exact and non-perturbative. If $\mu\tan\phi\gg1$, then  $m^z\approx \sign(\mu)\tilde g_L$ so the magnetization does not depend on the strength of the local field, but only the properties of the chain Hamiltonian via the term $\tilde g_L$. We found that when $|\lambda|<1$ it is $\Xi\approx e^{L/\alpha}$, $\alpha>0$ so in this region the approximation $m^z\approx \sign(\mu)\tilde g_L$ is well justified, as long as $\mu$ is not exponentially small in the system size.

\subsection{Comparison with numerical results}
\begin{figure}[t]
  \centering
  \includegraphics[width=0.45\textwidth]{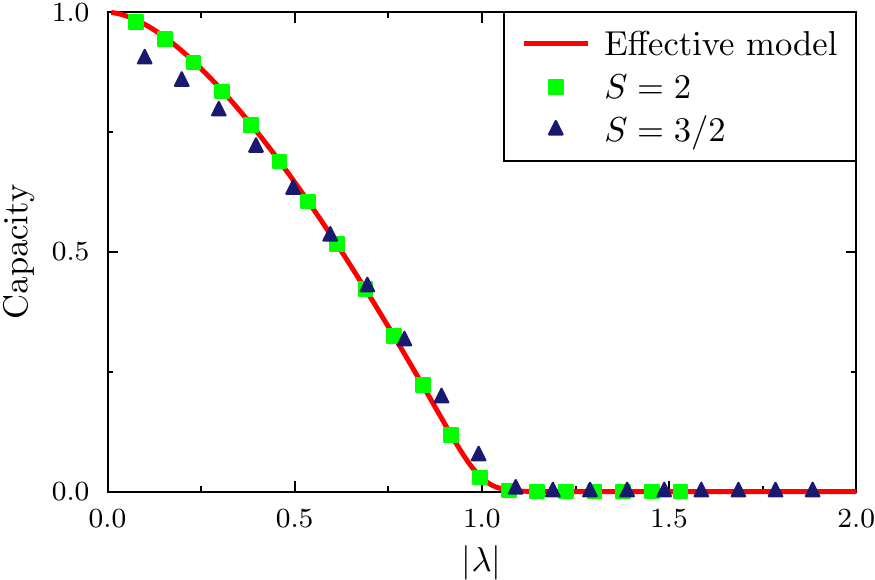}
  \caption{Capacity as a function of $\lambda$ as in Fig.~3 of the main 
  text. We keep $D=-7E$ with $B_x=0$ for $S=2$ and $B_x=E$ for $S=3/2$.
  The DMRG simulations are done for different values of $D$, and changing
  $E$ and $B_x$ such that the ratios $E/D$ and $B_x/D$ is fixed. The value 
  of $\lambda$ is then obtained from the equations of Table I in the 
  main text.
  }
  \label{fig:validity}
\end{figure}

In Fig.~\ref{fig:validity} we can 
extend the analysis performed in Fig.~3 of the 
main text to study the validity of the effective model for larger values
of $D$ while maintaining a fixed ratio $D/E$, 
to see whether one can observe the predicted phase transition also in
spin 2 and spin 3/2 systems. 
In principle, 
by enlarging $D$ with the typical experimental constraint that the ratio
$D/E$ is fixed, one may break at a certain point the assumption
of the effective theory, namely that $D$ is the only large parameter. 
Nonetheless, as seen in  Fig.~\ref{fig:validity}, the spin 2
simulations are in excellent  agreement 
with the theoretical predictions of the 
effective model, while for $S=3/2$ there is a small deviation, possibly 
due to higher order terms in the effective Hamiltonian that have 
been neglected. Therefore, as predicted by the effective model, one can
generally observe a transition for all $S\ge1$. 
However, only for $S=1$ this 
transitions can be obtained for reasonable values of $D,E,J$ and $B_x$. 
For instance, for $D=-7E$ the effective $\lambda$ for $S=2$ is 
$\lambda\simeq 3E^2/(8|D|J) \simeq 3|D|/400J$ so that the critical value 
$\lambda=1$ is obtained for the very large value $D\simeq -133 J$,
while for $S=3/2$ is obtained for $D\simeq-81 J$. This has to be compared 
with the much lower transition value $D=-7J$ observed in Fig.~3 of the main
text for $S=1$ systems. 

\begin{figure}[t]
  \centering
  \includegraphics[width=0.45\textwidth]{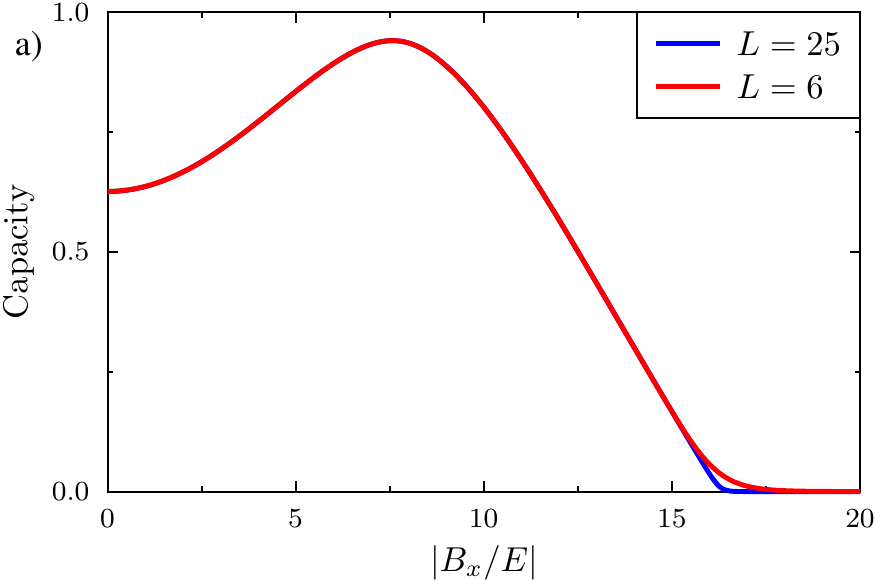}
  \includegraphics[width=0.45\textwidth]{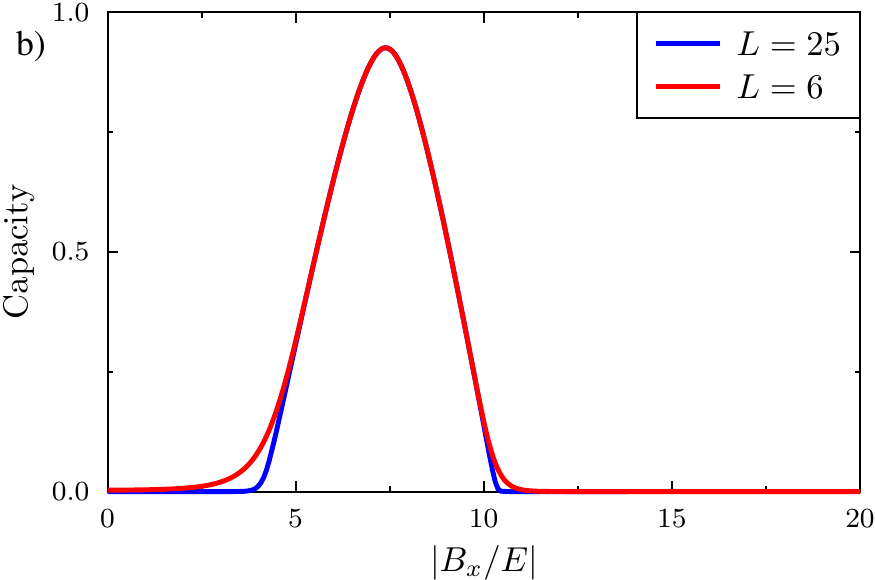}
  \caption{ Comparison between the capacity obtained with DMRG for $L=6$ and 
    $L=25$. The parameters are the same of Figs. 4a and 4b of the main text. 
  }
  \label{fig:check}
\end{figure}

Finally in Fig.~\ref{fig:check} we study the length dependence in the 
capacity. As predicted by Eq.~\eqref{e.mzbeta}, finite size effects are 
negligible (exponentially small) far from the critical points. 
On the other hand, around $\lambda\approx1$ the capacity displays 
a length dependence because of the critical scaling of the magnetization.

\section{Classical capacity of a classical Ising chain}
\label{a:classical}
A classical Ising chain defines a binary symmetric channel where the bit-flip probability $P_f$ is related to the average magnetization $m_j^z=\mean{\tau_j^z}$ by the relation $m_L^z=\pm[(1-P_f)-P_f]$, where the sign depends on the state of the island. Indeed, the output magnetization is the difference between the probability of having positive magnetization and that of having a negative one.  Therefore
\begin{align}
  C&=1-H_2\left[\frac{1+|m^z_L|}2\right]~,
  \label{e.capDlarg}
\end{align}
where $m_z=(-1)^L\tanh(\beta\mu)\tanh^{L-1}(\beta J)$ and $|m_z|\stackrel{L\gg 1}{\simeq}\exp(-2 L e^{-2\beta J})$. 
At zero temperature, $P_f=0$ and the chain perfectly transfers information; at higher temperature the capacity goes down as $C\simeq \exp(-4 L e^{-2\beta J})$, where $J$ is the strength of the Ising coupling. Therefore, thermal fluctuations limits the maximum length of the chain for reliable information transfer to $L\approx e^{2\beta J}$.

\section{Quantum encoding and decoding}
\label{a:quantum}
Quantum systems enable new forms of information processing which exploit peculiar quantum features such as superposition and entanglement. Indeed the quantum (spin) states are not described anymore by a binary value, but rather by a $(2S{+}1){\times}(2S{+}1)$ density matrix which encodes much more information. 
In our calculation for simplicity we used the 
capacity of a {\it classical} channel: even though our channel is physically implemented with quantum objects, both our inputs and outputs are binary values.  
On the other hand, in the general formalism, both input and output states are described by a quantum density matrix. 
To send classical information, namely a collection of binary inputs, the two input states, 0 and 1, are encoded into some quantum input states $\psi_0$ and $\psi_1$ that are sent through the channel. The resulting output state at the end of the channel (namely at the last spin of the chain) is then described by  a density matrix $\rho(\psi)$ that depends on the input state $\psi$. From the reconstructed states $\rho(\psi)$ for different $\psi$ one can then define the classical capacity of a quantum channel $C_Q$, namely the capacity of a quantum channel to transfer classical information. 
Considering only a single use of the channel, the capacity $C_Q$ is defined 
as
\begin{align}
  C_Q = \max_{p_i,\psi_i} S\left[\sum_i p_i \rho(\psi_i)\right] - 
  \sum_i p_i S\left[\rho(\psi_i)\right]~,
  \label{e.classcap}
\end{align}
where $S[\rho]=-\Tr\left[\rho\log\rho\right]$. When multiple uses of the channel are considered, then $C_Q$ provides a lower bound to the capacity \cite{giovannetti2005information}. The capacity $C_Q$ is more complicated to measure than Eq. \eqref{e.capacity} because  requires a full state tomography of the output state to determine all of the components of the density matrix. Moreover, it requires also an optimization over the initial encoding of the bits into two states $\psi$. 

In order to compute $C_Q$ and to compare it with $C$ we make some simplifications. First we assume that $\psi_0$ and $\psi_1$, namely the states of the island, can only be initialized in either up or down along the $z$ direction, as discussed in the main text. Secondly, we consider the effective description of Table \ref{tab:hameff}. In this effective spin-1/2 system, the final state is completely specified only by $m_L^x$, $m_L^z$, as we found that $m_L^y$ is always zero. On the other hand, for a general $S>1/2$ spin model the density matrix contains more elements and is not specified solely by the magnetization. Moreover, if $(m_L^x, 0, m_L^z)$ is the magnetic configuration when the island has positive magnetization, then $(m_L^x, 0, -m_L^z)$ is the configuration when the island is down. With these assumptions we find the capacity $C_Q(m_L^x, m_L^z) $ as a function of the two magnetizations
\begin{align*}
C_Q(m_L^x, m_L^z) &= C_m\left(\sqrt{(m_L^x)^2+(m_L^z)^2}\right) - C_m(m_L^x)~,\\
C(m) &= 1-H_2\left[\frac{1+m}2\right]~.
\end{align*}
It is  $C_Q(0,m_L^z) = C$ where $C\equiv C_m(m_L^z)$ is the capacity described in the main text  and in 
\eqref{e.capDlarg}, $C_Q(m_L^x,0) = 0$ and in general $C_Q(m_L^x,m_L^z) \ge C$. Therefore, when $0<C<1$, a quantum state encoding and decoding can enhance the capacity of the channel, but it cannot turn a broken channel into a non-broken one ($C_Q=0$ whenever $C=0$).

\section{Appearance of an effective field for $S=1$}

 The appearance of the effective transverse field $\lambda$ can be explained in terms of the level structure. For spin 1 systems, when $E=B=0$ the states $\ket{S^z{=}{{\pm}}1}$ are degenerate while $\ket{S^z{=}0}$ has a higher energy (the difference being $\simeq |D|$). The transverse anisotropy couples the states $\ket{S^z{=}{{\pm}}1}$ with an energy $\approx E$ without affecting $\ket{S^z{=}0}$. On the other hand, a field in the $xy$-plane 
enables an effective coupling $\simeq B/D$ between the states $\ket{S^z{=}{{\pm}}1}$ mediated by a two-step transition through  the virtual population of $\ket{S^z{=}0}$. In the limit $|D|\gg E,B$, this results in an effective global planar field shown in Table \ref{tab:hameff}.

\section{Ferromagnetic coupling}
\begin{figure*}[h]
  \centering
  \includegraphics[width=0.9\textwidth]{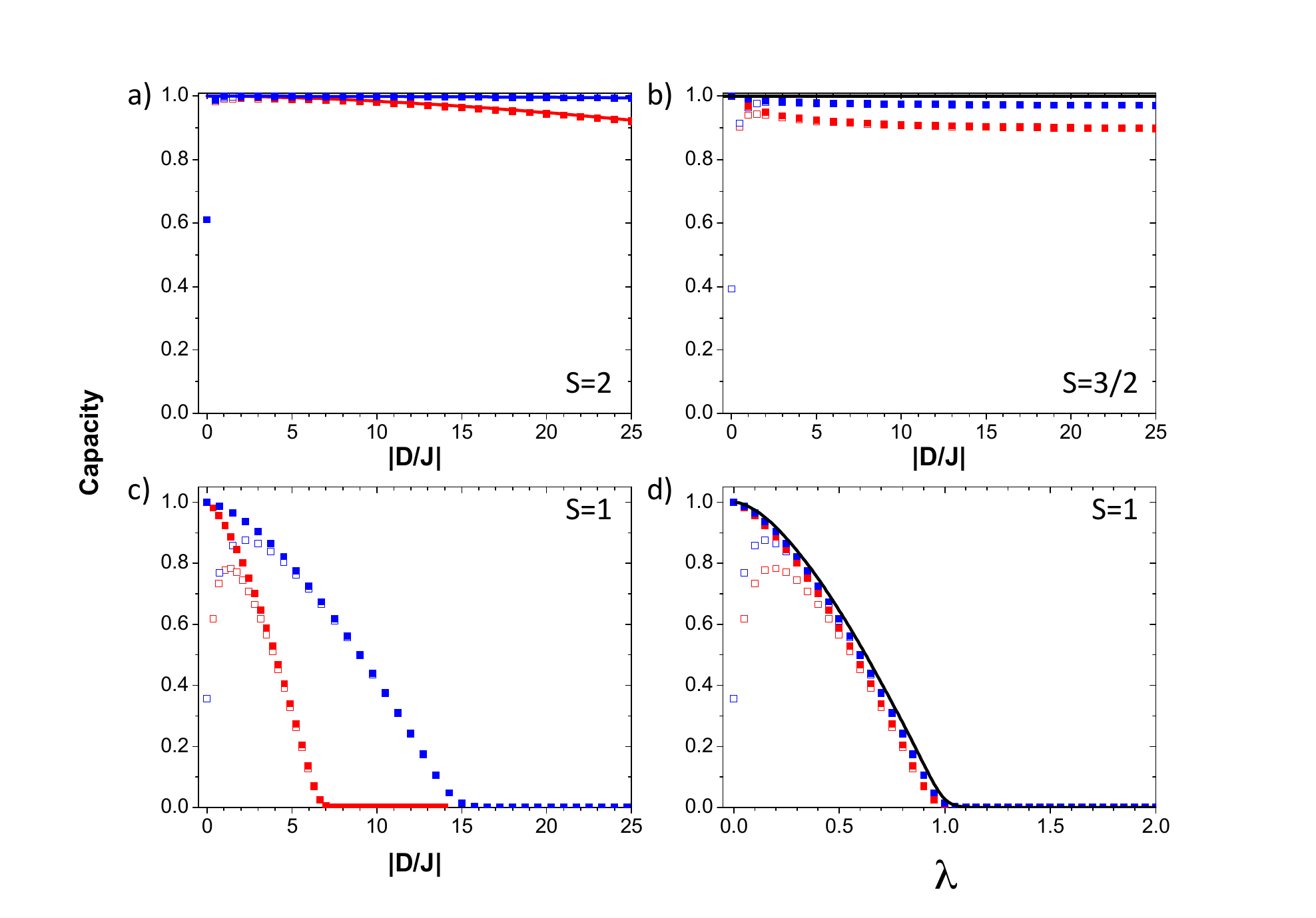}
  \caption{
  { Impact of quantum fluctuations on spin chain capacities for ferromagnetic and antiferromagnetic coupling.} a) Capacity of an $S=2$ spin chain with $L=25$, $B_0=J$, and $B_x=0$, as obtained from DMRG calculations, plotted as a function of $|D/J|$ for $D=-7E$ (blue squares) and $D=-15E$ (red squares). Open squares are for antiferromagnetic coupling while solid squares are for ferromagnetic coupling. The ratio $D/E$ is kept fixed as $|D/J|$ increases. Solid lines show the results obtained from the effective model with the same parameters. b) Same as panel (a) with $S=3/2$. c) Same as panel (a) with $S=1$. d) Same as panel (c) but plotted again $\lambda$. Solid line shows the results obtained from the effective model.}   
  \label{fig:fluctuationsFMAFM}
\end{figure*}

\begin{figure*}[h]
  \centering
  \includegraphics[width=.6\textwidth]{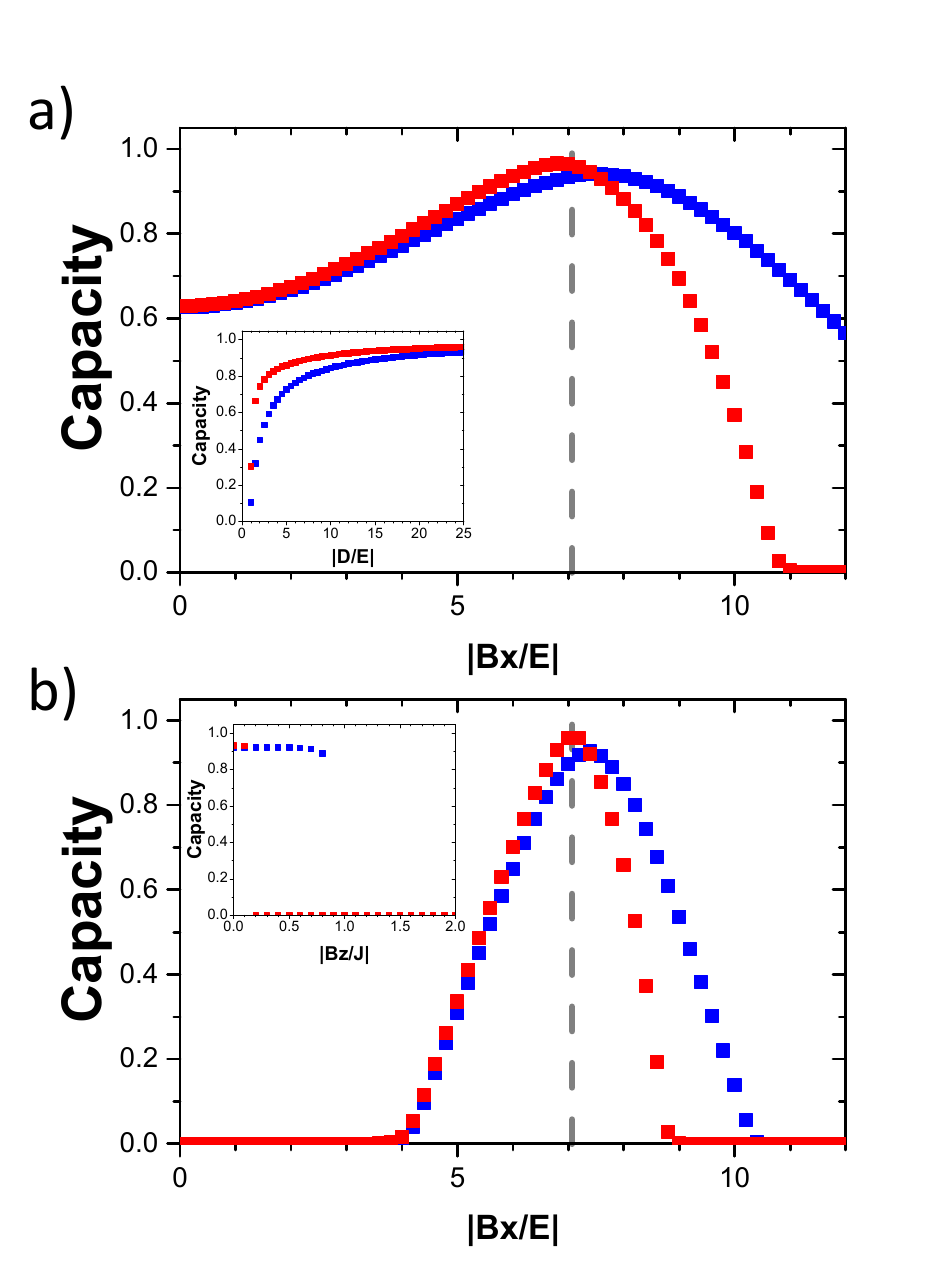}
  \caption{
  { Gating capacity via quantum fluctuations with an externally applied magnetic field.} 
   a) Capacity of an $S=1$ spin chain with with $L=25$, $B_0=J$, $D=-25E$, and $E=0.5J$, as obtained from DMRG simulations, plotted as a function of $|B_x/J|$ for antiferromagnetic (blue) and ferromagnetic (red) coupling. A maximum is observed close to $B_x=B_c=\sqrt{-2DE}$ (vertical dashed gray line). Inset shows the capacity as a function of $|D/E|$ when $B_x = B_c$. Same as panel (a) with $E=1.5J$, so that quantum fluctuations completely destroy the capacity at $B=0$. Inset shows the capacity as a function of $B_z$ when $B_x = B_c$.}
  \label{fig:fieldFMAFM}
\end{figure*}

In the main text, we focus on antiferromagnetic (AFM) Heisenberg coupling between the spins of an atomic chain. However, since ferromagnetic (FM) coupling is also possible in atomic systems \cite{spinelli2014imaging}, we consider those results as well. For DMRG calculations, we simply use $J<0$ rather than $J>0$; in the effective model shown in Table \ref{tab:hameff}, $\mathcal H_{\rm eff} \rightarrow  \mu \tau_1^z + \sum_n\left(\lambda \tau_n^x - \tau_n^z\tau_{n+1}^z\right) $. As seen in Supplementary Fig. \ref{fig:fluctuationsFMAFM}, small differences in the capacity are seen between FM and AFM chains when $D$ is small, though these become negligible as $D$ increases. Additional small deviations in the capacity are seen between FM and AFM chains in the presence of a magnetic field (Supplementary Fig. \ref{fig:fieldFMAFM}), though the qualitative behaviour remains the same in both cases. Furthermore, although FM systems generally have a higher capacity than their AFM counterparts, this capacity is highly sensitive to external fields in the $z$-direction. The effects of this sensitivity are shown in the inset of Supplementary Fig. \ref{fig:fieldFMAFM}b, where it is clear that a small field $B_z$ overcomes the effect of the magnetic islands for FM systems, while AFM chains are more stable against this perturbation.

\section{Capacity and ground state degeneracy}
\begin{figure*}[h]
  \centering
  \includegraphics[width=1.0\textwidth]{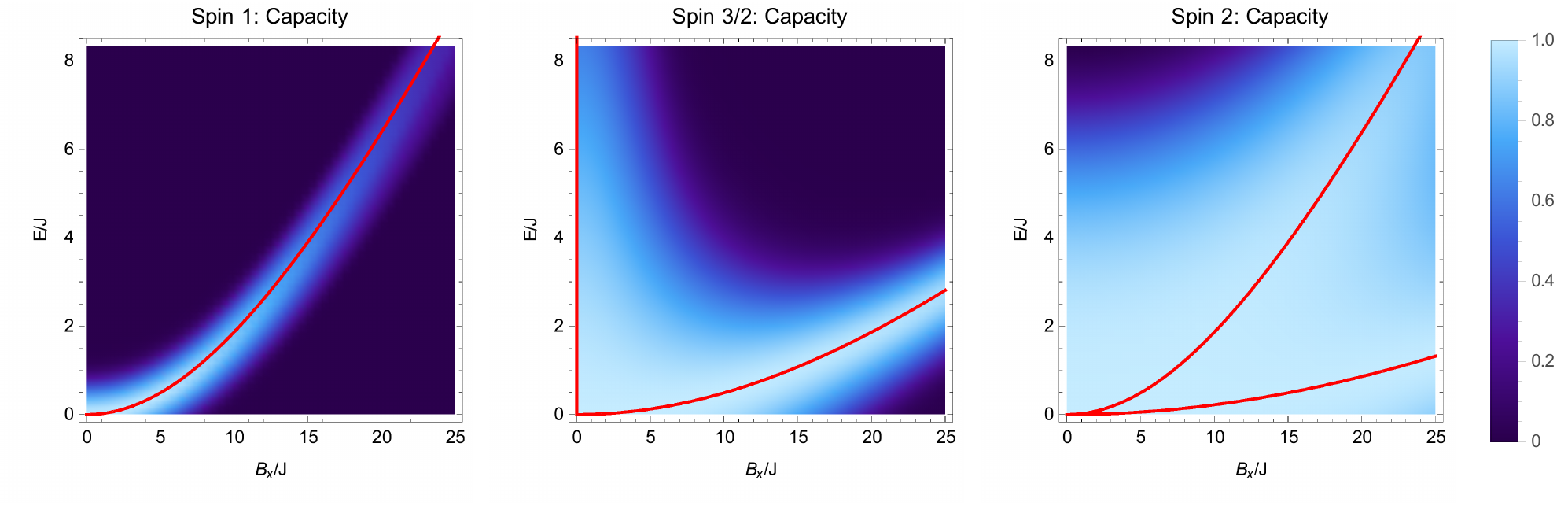}
  \includegraphics[width=1.0\textwidth]{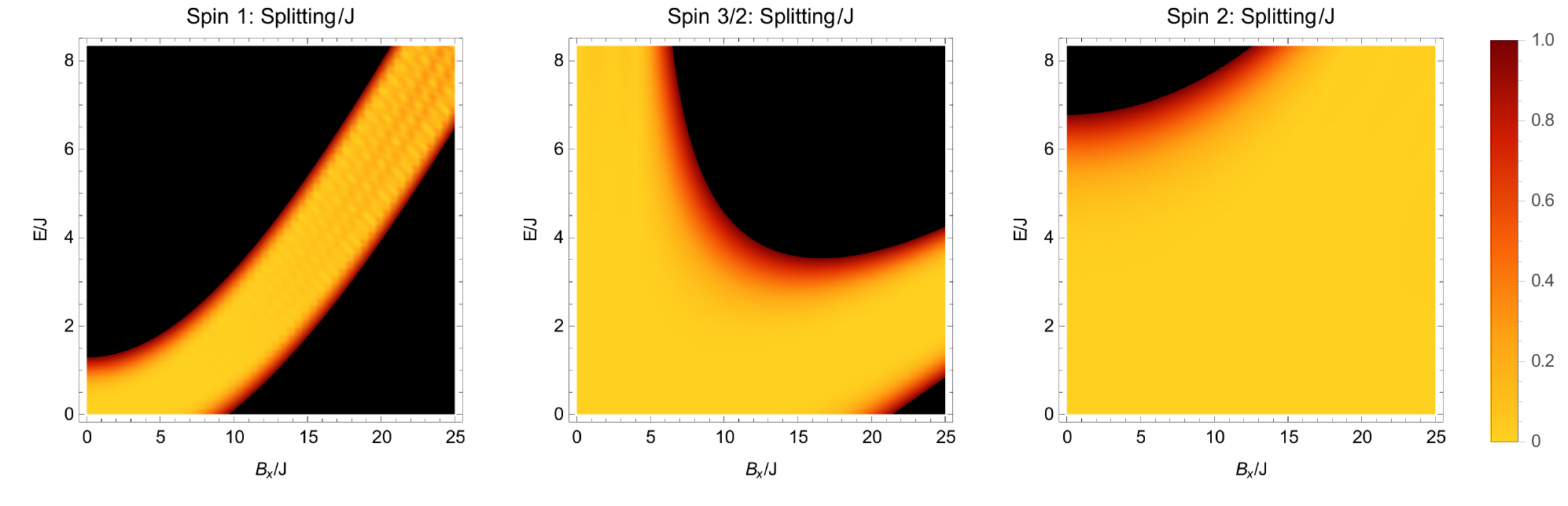}
  \caption{
  { Relationship between high-capacity and ground state degeneracy.} 
  Capacity and energy splitting between the ground state and the first excited state for different spins, $L=6$ (spin 1), $L=5$ (spin 3/2), $L=4$ (spin 2), $D=-25J$ and $B_0=J$ (for the capacity), $B_0=0$ (for the splitting). 
  Plots are shown for different values of $B_x\in[0,|D|]$ and $E\in[0,|D|/3]$. 
  The red line in the capacity marks the curves of zero SS-QST splitting. 
}
  \label{fig:CapVsGap}
\end{figure*}

As discussed via the effective Ising model, maximal capacity corresponds to $\lambda=0$ where the Ising chain has a degenerate ground state. The appearance of such degeneracy can be understood (at least to the leading order in the $J/|D|$ expansion) at the single spin level in terms of SS-QST and diabolic points \cite{wernsdorfer1999quantum}. Indeed, the single-spin curves which define the ``diabolic points'' almost completely match the curves of maximal capacity, as shown in Fig.~\ref{fig:CapVsGap}. Such relationship between high-capacity and (almost) degenerate ground states is extended to the many-body case in Fig. \ref{fig:CapVsGap} where one can see that the regions of high-capacity are in one-to-one correspondence to the regions where the energy difference between the many-body ground state and the first excited state is small. 

To explain this fact let us consider again the Ising effective description. As shown in the main text, non-zero capacity corresponds to the ordered phase $|\lambda|<1$. This phase in the thermodynamic limit is characterized by two degenerate ground states, while for finite chains there is a finite energy splitting between a unique ground state and the first excited state which goes to zero in the thermodynamic limit. Therefore, unlike the single-spin case where degeneracy occurs only at specific discrete values of $B_x$, in a chain the degeneracy occurs in a whole region. 
For instance in spin 1 chains the ordered region $-1<\lambda<1$ corresponds to $\sqrt{2|D|(E-J)}<B_x<\sqrt{2|D|(E+J)}$, when $E>J$, and so it can be expanded with a larger $J$. 

Moreover, as shown in Fig.~\ref{fig:CapVsGap}, this relationship between quasi-degeneracy and high-capacity is more general than the effective Ising description as it holds also when $B_x$ and $E/3$ are comparable with $|D|$ where the perturbative effective Ising model becomes less accurate.

\begin{thebibliography}{44}%
\makeatletter
\providecommand \@ifxundefined [1]{%
 \@ifx{#1\undefined}
}%
\providecommand \@ifnum [1]{%
 \ifnum #1\expandafter \@firstoftwo
 \else \expandafter \@secondoftwo
 \fi
}%
\providecommand \@ifx [1]{%
 \ifx #1\expandafter \@firstoftwo
 \else \expandafter \@secondoftwo
 \fi
}%
\providecommand \natexlab [1]{#1}%
\providecommand \enquote  [1]{``#1''}%
\providecommand \bibnamefont  [1]{#1}%
\providecommand \bibfnamefont [1]{#1}%
\providecommand \citenamefont [1]{#1}%
\providecommand \href@noop [0]{\@secondoftwo}%
\providecommand \href [0]{\begingroup \@sanitize@url \@href}%
\providecommand \@href[1]{\@@startlink{#1}\@@href}%
\providecommand \@@href[1]{\endgroup#1\@@endlink}%
\providecommand \@sanitize@url [0]{\catcode `\\12\catcode `\$12\catcode
  `\&12\catcode `\#12\catcode `\^12\catcode `\_12\catcode `\%12\relax}%
\providecommand \@@startlink[1]{}%
\providecommand \@@endlink[0]{}%
\providecommand \url  [0]{\begingroup\@sanitize@url \@url }%
\providecommand \@url [1]{\endgroup\@href {#1}{\urlprefix }}%
\providecommand \urlprefix  [0]{URL }%
\providecommand \Eprint [0]{\href }%
\providecommand \doibase [0]{http://dx.doi.org/}%
\providecommand \selectlanguage [0]{\@gobble}%
\providecommand \bibinfo  [0]{\@secondoftwo}%
\providecommand \bibfield  [0]{\@secondoftwo}%
\providecommand \translation [1]{[#1]}%
\providecommand \BibitemOpen [0]{}%
\providecommand \bibitemStop [0]{}%
\providecommand \bibitemNoStop [0]{.\EOS\space}%
\providecommand \EOS [0]{\spacefactor3000\relax}%
\providecommand \BibitemShut  [1]{\csname bibitem#1\endcsname}%
\let\auto@bib@innerbib\@empty
\bibitem [{\citenamefont {Nielsen}\ and\ \citenamefont
  {Chuang}(2010)}]{nielsen2010quantum}%
  \BibitemOpen
  \bibfield  {author} {\bibinfo {author} {\bibfnamefont {M.~A.}\ \bibnamefont
  {Nielsen}}\ and\ \bibinfo {author} {\bibfnamefont {I.~L.}\ \bibnamefont
  {Chuang}},\ }\href@noop {} {\emph {\bibinfo {title} {Quantum computation and
  quantum information}}}\ (\bibinfo  {publisher} {Cambridge university press},\
  \bibinfo {year} {2010})\BibitemShut {NoStop}%
\bibitem [{\citenamefont {Devoret}(1995)}]{devoret1995quantum}%
  \BibitemOpen
  \bibfield  {author} {\bibinfo {author} {\bibfnamefont {M.~H.}\ \bibnamefont
  {Devoret}},\ }\href@noop {} {\bibfield  {journal} {\bibinfo  {journal} {Les
  Houches, Session LXIII}\ }\textbf {\bibinfo {volume} {7}} (\bibinfo {year}
  {1995})}\BibitemShut {NoStop}%
\bibitem [{\citenamefont {Wong}(2002)}]{wong2002beyond}%
  \BibitemOpen
  \bibfield  {author} {\bibinfo {author} {\bibfnamefont {H.-S.}\ \bibnamefont
  {Wong}},\ }\href@noop {} {\bibfield  {journal} {\bibinfo  {journal} {IBM
  Journal of Research and Development}\ }\textbf {\bibinfo {volume} {46}},\
  \bibinfo {pages} {133} (\bibinfo {year} {2002})}\BibitemShut {NoStop}%
\bibitem [{\citenamefont {Gambardella}\ \emph {et~al.}(2002)\citenamefont
  {Gambardella}, \citenamefont {Dallmeyer}, \citenamefont {Maiti},
  \citenamefont {Malagoli}, \citenamefont {Eberhardt}, \citenamefont {Kern},\
  and\ \citenamefont {Carbone}}]{gambardella2002ferromagnetism}%
  \BibitemOpen
  \bibfield  {author} {\bibinfo {author} {\bibfnamefont {P.}~\bibnamefont
  {Gambardella}}, \bibinfo {author} {\bibfnamefont {A.}~\bibnamefont
  {Dallmeyer}}, \bibinfo {author} {\bibfnamefont {K.}~\bibnamefont {Maiti}},
  \bibinfo {author} {\bibfnamefont {M.}~\bibnamefont {Malagoli}}, \bibinfo
  {author} {\bibfnamefont {W.}~\bibnamefont {Eberhardt}}, \bibinfo {author}
  {\bibfnamefont {K.}~\bibnamefont {Kern}}, \ and\ \bibinfo {author}
  {\bibfnamefont {C.}~\bibnamefont {Carbone}},\ }\href@noop {} {\bibfield
  {journal} {\bibinfo  {journal} {Nature}\ }\textbf {\bibinfo {volume} {416}},\
  \bibinfo {pages} {301} (\bibinfo {year} {2002})}\BibitemShut {NoStop}%
\bibitem [{\citenamefont {Lee}\ \emph {et~al.}(2004)\citenamefont {Lee},
  \citenamefont {Ho},\ and\ \citenamefont {Persson}}]{lee2004spin}%
  \BibitemOpen
  \bibfield  {author} {\bibinfo {author} {\bibfnamefont {H.}~\bibnamefont
  {Lee}}, \bibinfo {author} {\bibfnamefont {W.}~\bibnamefont {Ho}}, \ and\
  \bibinfo {author} {\bibfnamefont {M.}~\bibnamefont {Persson}},\ }\href@noop
  {} {\bibfield  {journal} {\bibinfo  {journal} {Phys. Rev. Lett.}\ }\textbf
  {\bibinfo {volume} {92}},\ \bibinfo {pages} {186802} (\bibinfo {year}
  {2004})}\BibitemShut {NoStop}%
\bibitem [{\citenamefont {Hirjibehedin}\ \emph {et~al.}(2006)\citenamefont
  {Hirjibehedin}, \citenamefont {Lutz},\ and\ \citenamefont
  {Heinrich}}]{hirjibehedin2006spin}%
  \BibitemOpen
  \bibfield  {author} {\bibinfo {author} {\bibfnamefont {C.~F.}\ \bibnamefont
  {Hirjibehedin}}, \bibinfo {author} {\bibfnamefont {C.~P.}\ \bibnamefont
  {Lutz}}, \ and\ \bibinfo {author} {\bibfnamefont {A.~J.}\ \bibnamefont
  {Heinrich}},\ }\href@noop {} {\bibfield  {journal} {\bibinfo  {journal}
  {Science}\ }\textbf {\bibinfo {volume} {312}},\ \bibinfo {pages} {1021}
  (\bibinfo {year} {2006})}\BibitemShut {NoStop}%
\bibitem [{\citenamefont {Kitchen}\ \emph {et~al.}(2006)\citenamefont
  {Kitchen}, \citenamefont {Richardella}, \citenamefont {Tang}, \citenamefont
  {Flatt{\'e}},\ and\ \citenamefont {Yazdani}}]{kitchen2006atom}%
  \BibitemOpen
  \bibfield  {author} {\bibinfo {author} {\bibfnamefont {D.}~\bibnamefont
  {Kitchen}}, \bibinfo {author} {\bibfnamefont {A.}~\bibnamefont
  {Richardella}}, \bibinfo {author} {\bibfnamefont {J.-M.}\ \bibnamefont
  {Tang}}, \bibinfo {author} {\bibfnamefont {M.~E.}\ \bibnamefont
  {Flatt{\'e}}}, \ and\ \bibinfo {author} {\bibfnamefont {A.}~\bibnamefont
  {Yazdani}},\ }\href@noop {} {\bibfield  {journal} {\bibinfo  {journal}
  {Nature}\ }\textbf {\bibinfo {volume} {442}},\ \bibinfo {pages} {436}
  (\bibinfo {year} {2006})}\BibitemShut {NoStop}%
\bibitem [{\citenamefont {Chen}\ \emph {et~al.}(2008)\citenamefont {Chen},
  \citenamefont {Fu}, \citenamefont {Ji}, \citenamefont {Zhang}, \citenamefont
  {Cheng}, \citenamefont {Ma}, \citenamefont {Zou}, \citenamefont {Duan},
  \citenamefont {Jia},\ and\ \citenamefont {Xue}}]{chen2008probing}%
  \BibitemOpen
  \bibfield  {author} {\bibinfo {author} {\bibfnamefont {X.}~\bibnamefont
  {Chen}}, \bibinfo {author} {\bibfnamefont {Y.-S.}\ \bibnamefont {Fu}},
  \bibinfo {author} {\bibfnamefont {S.-H.}\ \bibnamefont {Ji}}, \bibinfo
  {author} {\bibfnamefont {T.}~\bibnamefont {Zhang}}, \bibinfo {author}
  {\bibfnamefont {P.}~\bibnamefont {Cheng}}, \bibinfo {author} {\bibfnamefont
  {X.-C.}\ \bibnamefont {Ma}}, \bibinfo {author} {\bibfnamefont {X.-L.}\
  \bibnamefont {Zou}}, \bibinfo {author} {\bibfnamefont {W.-H.}\ \bibnamefont
  {Duan}}, \bibinfo {author} {\bibfnamefont {J.-F.}\ \bibnamefont {Jia}}, \
  and\ \bibinfo {author} {\bibfnamefont {Q.-K.}\ \bibnamefont {Xue}},\
  }\href@noop {} {\bibfield  {journal} {\bibinfo  {journal} {Phys. Rev. Lett.}\
  }\textbf {\bibinfo {volume} {101}},\ \bibinfo {pages} {197208} (\bibinfo
  {year} {2008})}\BibitemShut {NoStop}%
\bibitem [{\citenamefont {Khajetoorians}\ \emph {et~al.}(2011)\citenamefont
  {Khajetoorians}, \citenamefont {Wiebe}, \citenamefont {Chilian},\ and\
  \citenamefont {Wiesendanger}}]{khajetoorians2011realizing}%
  \BibitemOpen
  \bibfield  {author} {\bibinfo {author} {\bibfnamefont {A.~A.}\ \bibnamefont
  {Khajetoorians}}, \bibinfo {author} {\bibfnamefont {J.}~\bibnamefont
  {Wiebe}}, \bibinfo {author} {\bibfnamefont {B.}~\bibnamefont {Chilian}}, \
  and\ \bibinfo {author} {\bibfnamefont {R.}~\bibnamefont {Wiesendanger}},\
  }\href@noop {} {\bibfield  {journal} {\bibinfo  {journal} {Science}\ }\textbf
  {\bibinfo {volume} {332}},\ \bibinfo {pages} {1062} (\bibinfo {year}
  {2011})}\BibitemShut {NoStop}%
\bibitem [{\citenamefont {Khajetoorians}\ \emph {et~al.}(2012)\citenamefont
  {Khajetoorians}, \citenamefont {Wiebe}, \citenamefont {Chilian},
  \citenamefont {Lounis}, \citenamefont {Bl{\"u}gel},\ and\ \citenamefont
  {Wiesendanger}}]{khajetoorians2012atom}%
  \BibitemOpen
  \bibfield  {author} {\bibinfo {author} {\bibfnamefont {A.~A.}\ \bibnamefont
  {Khajetoorians}}, \bibinfo {author} {\bibfnamefont {J.}~\bibnamefont
  {Wiebe}}, \bibinfo {author} {\bibfnamefont {B.}~\bibnamefont {Chilian}},
  \bibinfo {author} {\bibfnamefont {S.}~\bibnamefont {Lounis}}, \bibinfo
  {author} {\bibfnamefont {S.}~\bibnamefont {Bl{\"u}gel}}, \ and\ \bibinfo
  {author} {\bibfnamefont {R.}~\bibnamefont {Wiesendanger}},\ }\href@noop {}
  {\bibfield  {journal} {\bibinfo  {journal} {Nature Physics}\ }\textbf
  {\bibinfo {volume} {8}},\ \bibinfo {pages} {497} (\bibinfo {year}
  {2012})}\BibitemShut {NoStop}%
\bibitem [{\citenamefont {Loth}\ \emph {et~al.}(2012)\citenamefont {Loth},
  \citenamefont {Baumann}, \citenamefont {Lutz}, \citenamefont {Eigler},\ and\
  \citenamefont {Heinrich}}]{loth2012bistability}%
  \BibitemOpen
  \bibfield  {author} {\bibinfo {author} {\bibfnamefont {S.}~\bibnamefont
  {Loth}}, \bibinfo {author} {\bibfnamefont {S.}~\bibnamefont {Baumann}},
  \bibinfo {author} {\bibfnamefont {C.~P.}\ \bibnamefont {Lutz}}, \bibinfo
  {author} {\bibfnamefont {D.}~\bibnamefont {Eigler}}, \ and\ \bibinfo {author}
  {\bibfnamefont {A.~J.}\ \bibnamefont {Heinrich}},\ }\href@noop {} {\bibfield
  {journal} {\bibinfo  {journal} {Science}\ }\textbf {\bibinfo {volume}
  {335}},\ \bibinfo {pages} {196} (\bibinfo {year} {2012})}\BibitemShut
  {NoStop}%
\bibitem [{\citenamefont {Spinelli}\ \emph {et~al.}(2014)\citenamefont
  {Spinelli}, \citenamefont {Bryant}, \citenamefont {Delgado}, \citenamefont
  {Fern{\'a}ndez-Rossier},\ and\ \citenamefont {Otte}}]{spinelli2014imaging}%
  \BibitemOpen
  \bibfield  {author} {\bibinfo {author} {\bibfnamefont {A.}~\bibnamefont
  {Spinelli}}, \bibinfo {author} {\bibfnamefont {B.}~\bibnamefont {Bryant}},
  \bibinfo {author} {\bibfnamefont {F.}~\bibnamefont {Delgado}}, \bibinfo
  {author} {\bibfnamefont {J.}~\bibnamefont {Fern{\'a}ndez-Rossier}}, \ and\
  \bibinfo {author} {\bibfnamefont {A.~F.}\ \bibnamefont {Otte}},\ }\href@noop
  {} {\bibfield  {journal} {\bibinfo  {journal} {Nature Materials}\ }\textbf
  {\bibinfo {volume} {13}},\ \bibinfo {pages} {782} (\bibinfo {year}
  {2014})}\BibitemShut {NoStop}%
\bibitem [{\citenamefont {Nadj-Perge}\ \emph {et~al.}(2014)\citenamefont
  {Nadj-Perge}, \citenamefont {Drozdov}, \citenamefont {Li}, \citenamefont
  {Chen}, \citenamefont {Jeon}, \citenamefont {Seo}, \citenamefont {MacDonald},
  \citenamefont {Bernevig},\ and\ \citenamefont
  {Yazdani}}]{nadj2014observation}%
  \BibitemOpen
  \bibfield  {author} {\bibinfo {author} {\bibfnamefont {S.}~\bibnamefont
  {Nadj-Perge}}, \bibinfo {author} {\bibfnamefont {I.~K.}\ \bibnamefont
  {Drozdov}}, \bibinfo {author} {\bibfnamefont {J.}~\bibnamefont {Li}},
  \bibinfo {author} {\bibfnamefont {H.}~\bibnamefont {Chen}}, \bibinfo {author}
  {\bibfnamefont {S.}~\bibnamefont {Jeon}}, \bibinfo {author} {\bibfnamefont
  {J.}~\bibnamefont {Seo}}, \bibinfo {author} {\bibfnamefont {A.~H.}\
  \bibnamefont {MacDonald}}, \bibinfo {author} {\bibfnamefont {B.~A.}\
  \bibnamefont {Bernevig}}, \ and\ \bibinfo {author} {\bibfnamefont
  {A.}~\bibnamefont {Yazdani}},\ }\href@noop {} {\bibfield  {journal} {\bibinfo
   {journal} {Science}\ }\textbf {\bibinfo {volume} {346}},\ \bibinfo {pages}
  {602} (\bibinfo {year} {2014})}\BibitemShut {NoStop}%
\bibitem [{\citenamefont {Weber}\ \emph {et~al.}(2012)\citenamefont {Weber},
  \citenamefont {Mahapatra}, \citenamefont {Ryu}, \citenamefont {Lee},
  \citenamefont {Fuhrer}, \citenamefont {Reusch}, \citenamefont {Thompson},
  \citenamefont {Lee}, \citenamefont {Klimeck}, \citenamefont {Hollenberg}
  \emph {et~al.}}]{weber2012ohm}%
  \BibitemOpen
  \bibfield  {author} {\bibinfo {author} {\bibfnamefont {B.}~\bibnamefont
  {Weber}}, \bibinfo {author} {\bibfnamefont {S.}~\bibnamefont {Mahapatra}},
  \bibinfo {author} {\bibfnamefont {H.}~\bibnamefont {Ryu}}, \bibinfo {author}
  {\bibfnamefont {S.}~\bibnamefont {Lee}}, \bibinfo {author} {\bibfnamefont
  {A.}~\bibnamefont {Fuhrer}}, \bibinfo {author} {\bibfnamefont
  {T.}~\bibnamefont {Reusch}}, \bibinfo {author} {\bibfnamefont
  {D.}~\bibnamefont {Thompson}}, \bibinfo {author} {\bibfnamefont
  {W.}~\bibnamefont {Lee}}, \bibinfo {author} {\bibfnamefont {G.}~\bibnamefont
  {Klimeck}}, \bibinfo {author} {\bibfnamefont {L.~C.}\ \bibnamefont
  {Hollenberg}},  \emph {et~al.},\ }\href@noop {} {\bibfield  {journal}
  {\bibinfo  {journal} {Science}\ }\textbf {\bibinfo {volume} {335}},\ \bibinfo
  {pages} {64} (\bibinfo {year} {2012})}\BibitemShut {NoStop}%
\bibitem [{\citenamefont {Menzel}\ \emph {et~al.}(2012)\citenamefont {Menzel},
  \citenamefont {Mokrousov}, \citenamefont {Wieser}, \citenamefont {Bickel},
  \citenamefont {Vedmedenko}, \citenamefont {Bl{\"u}gel}, \citenamefont
  {Heinze}, \citenamefont {von Bergmann}, \citenamefont {Kubetzka},\ and\
  \citenamefont {Wiesendanger}}]{menzel2012information}%
  \BibitemOpen
  \bibfield  {author} {\bibinfo {author} {\bibfnamefont {M.}~\bibnamefont
  {Menzel}}, \bibinfo {author} {\bibfnamefont {Y.}~\bibnamefont {Mokrousov}},
  \bibinfo {author} {\bibfnamefont {R.}~\bibnamefont {Wieser}}, \bibinfo
  {author} {\bibfnamefont {J.~E.}\ \bibnamefont {Bickel}}, \bibinfo {author}
  {\bibfnamefont {E.}~\bibnamefont {Vedmedenko}}, \bibinfo {author}
  {\bibfnamefont {S.}~\bibnamefont {Bl{\"u}gel}}, \bibinfo {author}
  {\bibfnamefont {S.}~\bibnamefont {Heinze}}, \bibinfo {author} {\bibfnamefont
  {K.}~\bibnamefont {von Bergmann}}, \bibinfo {author} {\bibfnamefont
  {A.}~\bibnamefont {Kubetzka}}, \ and\ \bibinfo {author} {\bibfnamefont
  {R.}~\bibnamefont {Wiesendanger}},\ }\href@noop {} {\bibfield  {journal}
  {\bibinfo  {journal} {Phys. Rev. Lett.}\ }\textbf {\bibinfo {volume} {108}},\
  \bibinfo {pages} {197204} (\bibinfo {year} {2012})}\BibitemShut {NoStop}%
\bibitem [{\citenamefont {Shannon}(1948)}]{shannon1948amathematical}%
  \BibitemOpen
  \bibfield  {author} {\bibinfo {author} {\bibfnamefont {C.}~\bibnamefont
  {Shannon}},\ }\href@noop {} {\bibfield  {journal} {\bibinfo  {journal} {Bell
  System Technical Journal}\ }\textbf {\bibinfo {volume} {27}},\ \bibinfo
  {pages} {379} (\bibinfo {year} {1948})}\BibitemShut {NoStop}%
\bibitem [{\citenamefont {Dutta}\ \emph {et~al.}(2015)\citenamefont {Dutta},
  \citenamefont {Aeppli}, \citenamefont {Chakrabarti}, \citenamefont
  {Divakaran}, \citenamefont {Rosenbaum},\ and\ \citenamefont
  {Sen}}]{dutta2015quantum}%
  \BibitemOpen
  \bibfield  {author} {\bibinfo {author} {\bibfnamefont {A.}~\bibnamefont
  {Dutta}}, \bibinfo {author} {\bibfnamefont {G.}~\bibnamefont {Aeppli}},
  \bibinfo {author} {\bibfnamefont {B.~K.}\ \bibnamefont {Chakrabarti}},
  \bibinfo {author} {\bibfnamefont {U.}~\bibnamefont {Divakaran}}, \bibinfo
  {author} {\bibfnamefont {T.~F.}\ \bibnamefont {Rosenbaum}}, \ and\ \bibinfo
  {author} {\bibfnamefont {D.}~\bibnamefont {Sen}},\ }\href@noop {} {\emph
  {\bibinfo {title} {Quantum Phase Transitions in Transverse Field Models}}}\
  (\bibinfo  {publisher} {Cambridge University Press},\ \bibinfo {year}
  {2015})\BibitemShut {NoStop}%
\bibitem [{\citenamefont {Sachdev}(2007)}]{sachdev2007quantum}%
  \BibitemOpen
  \bibfield  {author} {\bibinfo {author} {\bibfnamefont {S.}~\bibnamefont
  {Sachdev}},\ }\href@noop {} {\emph {\bibinfo {title} {Quantum phase
  transitions}}}\ (\bibinfo  {publisher} {Wiley Online Library},\ \bibinfo
  {year} {2007})\BibitemShut {NoStop}%
\bibitem [{\citenamefont {Wiesendanger}(2009)}]{wiesendanger2009SPSTMreview}%
  \BibitemOpen
  \bibfield  {author} {\bibinfo {author} {\bibfnamefont {R.}~\bibnamefont
  {Wiesendanger}},\ }\href@noop {} {\bibfield  {journal} {\bibinfo  {journal}
  {Rev. Mod. Phys.}\ }\textbf {\bibinfo {volume} {81}},\ \bibinfo {pages}
  {1495} (\bibinfo {year} {2009})}\BibitemShut {NoStop}%
\bibitem [{\citenamefont {Holevo}(1973)}]{holevo1973bounds}%
  \BibitemOpen
  \bibfield  {author} {\bibinfo {author} {\bibfnamefont {A.~S.}\ \bibnamefont
  {Holevo}},\ }\href@noop {} {\bibfield  {journal} {\bibinfo  {journal}
  {Problemy Peredachi Informatsii}\ }\textbf {\bibinfo {volume} {9}},\ \bibinfo
  {pages} {3} (\bibinfo {year} {1973})}\BibitemShut {NoStop}%
\bibitem [{\citenamefont {Giovannetti}\ and\ \citenamefont
  {Fazio}(2005)}]{giovannetti2005information}%
  \BibitemOpen
  \bibfield  {author} {\bibinfo {author} {\bibfnamefont {V.}~\bibnamefont
  {Giovannetti}}\ and\ \bibinfo {author} {\bibfnamefont {R.}~\bibnamefont
  {Fazio}},\ }\href@noop {} {\bibfield  {journal} {\bibinfo  {journal} {Phys.
  Rev. A}\ }\textbf {\bibinfo {volume} {71}},\ \bibinfo {pages} {032314}
  (\bibinfo {year} {2005})}\BibitemShut {NoStop}%
\bibitem [{\citenamefont {Giovannetti}\ \emph {et~al.}(2014)\citenamefont
  {Giovannetti}, \citenamefont {Garc{\'\i}a-Patr{\'o}n}, \citenamefont {Cerf},\
  and\ \citenamefont {Holevo}}]{giovannetti2014ultimate}%
  \BibitemOpen
  \bibfield  {author} {\bibinfo {author} {\bibfnamefont {V.}~\bibnamefont
  {Giovannetti}}, \bibinfo {author} {\bibfnamefont {R.}~\bibnamefont
  {Garc{\'\i}a-Patr{\'o}n}}, \bibinfo {author} {\bibfnamefont {N.}~\bibnamefont
  {Cerf}}, \ and\ \bibinfo {author} {\bibfnamefont {A.}~\bibnamefont
  {Holevo}},\ }\href@noop {} {\bibfield  {journal} {\bibinfo  {journal} {Nature
  Photonics}\ }\textbf {\bibinfo {volume} {8}},\ \bibinfo {pages} {796}
  (\bibinfo {year} {2014})}\BibitemShut {NoStop}%
\bibitem [{\citenamefont {Giovannetti}\ \emph {et~al.}(2013)\citenamefont
  {Giovannetti}, \citenamefont {Lloyd}, \citenamefont {Maccone},\ and\
  \citenamefont {Shapiro}}]{giovannetti2013electromagnetic}%
  \BibitemOpen
  \bibfield  {author} {\bibinfo {author} {\bibfnamefont {V.}~\bibnamefont
  {Giovannetti}}, \bibinfo {author} {\bibfnamefont {S.}~\bibnamefont {Lloyd}},
  \bibinfo {author} {\bibfnamefont {L.}~\bibnamefont {Maccone}}, \ and\
  \bibinfo {author} {\bibfnamefont {J.~H.}\ \bibnamefont {Shapiro}},\
  }\href@noop {} {\bibfield  {journal} {\bibinfo  {journal} {Nature Photonics}\
  }\textbf {\bibinfo {volume} {7}},\ \bibinfo {pages} {834} (\bibinfo {year}
  {2013})}\BibitemShut {NoStop}%
\bibitem [{\citenamefont {Banaszek}(2012)}]{banaszek2012quantum}%
  \BibitemOpen
  \bibfield  {author} {\bibinfo {author} {\bibfnamefont {K.}~\bibnamefont
  {Banaszek}},\ }\href@noop {} {\bibfield  {journal} {\bibinfo  {journal}
  {Nature Photonics}\ }\textbf {\bibinfo {volume} {6}},\ \bibinfo {pages} {351}
  (\bibinfo {year} {2012})}\BibitemShut {NoStop}%
\bibitem [{\citenamefont {Schollw{\"o}ck}(2011)}]{schollwock2011density}%
  \BibitemOpen
  \bibfield  {author} {\bibinfo {author} {\bibfnamefont {U.}~\bibnamefont
  {Schollw{\"o}ck}},\ }\href@noop {} {\bibfield  {journal} {\bibinfo  {journal}
  {Annals of Physics}\ }\textbf {\bibinfo {volume} {326}},\ \bibinfo {pages}
  {96} (\bibinfo {year} {2011})}\BibitemShut {NoStop}%
\bibitem [{\citenamefont {Delgado}\ and\ \citenamefont
  {Fern{\'a}ndez-Rossier}(2012)}]{delgado2012storage}%
  \BibitemOpen
  \bibfield  {author} {\bibinfo {author} {\bibfnamefont {F.}~\bibnamefont
  {Delgado}}\ and\ \bibinfo {author} {\bibfnamefont {J.}~\bibnamefont
  {Fern{\'a}ndez-Rossier}},\ }\href@noop {} {\bibfield  {journal} {\bibinfo
  {journal} {Phys. Rev. Lett.}\ }\textbf {\bibinfo {volume} {108}},\ \bibinfo
  {pages} {196602} (\bibinfo {year} {2012})}\BibitemShut {NoStop}%
\bibitem [{\citenamefont {Delgado}\ \emph {et~al.}(2015)\citenamefont
  {Delgado}, \citenamefont {Loth}, \citenamefont {Zielinski},\ and\
  \citenamefont {Fernández-Rossier}}]{delgado2014emergence}%
  \BibitemOpen
  \bibfield  {author} {\bibinfo {author} {\bibfnamefont {F.}~\bibnamefont
  {Delgado}}, \bibinfo {author} {\bibfnamefont {S.}~\bibnamefont {Loth}},
  \bibinfo {author} {\bibfnamefont {M.}~\bibnamefont {Zielinski}}, \ and\
  \bibinfo {author} {\bibfnamefont {J.}~\bibnamefont {Fernández-Rossier}},\
  }\href {http://stacks.iop.org/0295-5075/109/i=5/a=57001} {\bibfield
  {journal} {\bibinfo  {journal} {EPL (Europhysics Letters)}\ }\textbf
  {\bibinfo {volume} {109}},\ \bibinfo {pages} {57001} (\bibinfo {year}
  {2015})}\BibitemShut {NoStop}%
\bibitem [{\citenamefont {Jia}\ \emph {et~al.}(2015)\citenamefont {Jia},
  \citenamefont {Banchi}, \citenamefont {Bayat}, \citenamefont {Dong},\ and\
  \citenamefont {Bose}}]{jia2015integrated}%
  \BibitemOpen
  \bibfield  {author} {\bibinfo {author} {\bibfnamefont {N.}~\bibnamefont
  {Jia}}, \bibinfo {author} {\bibfnamefont {L.}~\bibnamefont {Banchi}},
  \bibinfo {author} {\bibfnamefont {A.}~\bibnamefont {Bayat}}, \bibinfo
  {author} {\bibfnamefont {G.}~\bibnamefont {Dong}}, \ and\ \bibinfo {author}
  {\bibfnamefont {S.}~\bibnamefont {Bose}},\ }\href@noop {} {\bibfield
  {journal} {\bibinfo  {journal} {Scientific Reports}\ }\textbf {\bibinfo
  {volume} {5}},\ \bibinfo {pages} {13665} (\bibinfo {year}
  {2015})}\BibitemShut {NoStop}%
\bibitem [{\citenamefont {Igl{\'o}i}\ and\ \citenamefont
  {Rieger}(1997)}]{igloi1997density}%
  \BibitemOpen
  \bibfield  {author} {\bibinfo {author} {\bibfnamefont {F.}~\bibnamefont
  {Igl{\'o}i}}\ and\ \bibinfo {author} {\bibfnamefont {H.}~\bibnamefont
  {Rieger}},\ }\href@noop {} {\bibfield  {journal} {\bibinfo  {journal}
  {Physical review letters}\ }\textbf {\bibinfo {volume} {78}},\ \bibinfo
  {pages} {2473} (\bibinfo {year} {1997})}\BibitemShut {NoStop}%
\bibitem [{\citenamefont {Burkhardt}\ and\ \citenamefont
  {Xue}(1991)}]{burkhardt1991density}%
  \BibitemOpen
  \bibfield  {author} {\bibinfo {author} {\bibfnamefont {T.~W.}\ \bibnamefont
  {Burkhardt}}\ and\ \bibinfo {author} {\bibfnamefont {T.}~\bibnamefont
  {Xue}},\ }\href@noop {} {\bibfield  {journal} {\bibinfo  {journal} {Physical
  review letters}\ }\textbf {\bibinfo {volume} {66}},\ \bibinfo {pages} {895}
  (\bibinfo {year} {1991})}\BibitemShut {NoStop}%
\bibitem [{\citenamefont {Garg}(1993)}]{garg1993topologically}%
  \BibitemOpen
  \bibfield  {author} {\bibinfo {author} {\bibfnamefont {A.}~\bibnamefont
  {Garg}},\ }\href@noop {} {\bibfield  {journal} {\bibinfo  {journal} {EPL
  (Europhysics Letters)}\ }\textbf {\bibinfo {volume} {22}},\ \bibinfo {pages}
  {205} (\bibinfo {year} {1993})}\BibitemShut {NoStop}%
\bibitem [{\citenamefont {Wernsdorfer}\ and\ \citenamefont
  {Sessoli}(1999)}]{wernsdorfer1999quantum}%
  \BibitemOpen
  \bibfield  {author} {\bibinfo {author} {\bibfnamefont {W.}~\bibnamefont
  {Wernsdorfer}}\ and\ \bibinfo {author} {\bibfnamefont {R.}~\bibnamefont
  {Sessoli}},\ }\href@noop {} {\bibfield  {journal} {\bibinfo  {journal}
  {Science}\ }\textbf {\bibinfo {volume} {284}},\ \bibinfo {pages} {133}
  (\bibinfo {year} {1999})}\BibitemShut {NoStop}%
\bibitem [{\citenamefont {Amico}\ \emph {et~al.}(2008)\citenamefont {Amico},
  \citenamefont {Fazio}, \citenamefont {Osterloh},\ and\ \citenamefont
  {Vedral}}]{amico2008entanglement}%
  \BibitemOpen
  \bibfield  {author} {\bibinfo {author} {\bibfnamefont {L.}~\bibnamefont
  {Amico}}, \bibinfo {author} {\bibfnamefont {R.}~\bibnamefont {Fazio}},
  \bibinfo {author} {\bibfnamefont {A.}~\bibnamefont {Osterloh}}, \ and\
  \bibinfo {author} {\bibfnamefont {V.}~\bibnamefont {Vedral}},\ }\href@noop {}
  {\bibfield  {journal} {\bibinfo  {journal} {Rev. Mod. Phys.}\ }\textbf
  {\bibinfo {volume} {80}},\ \bibinfo {pages} {517} (\bibinfo {year}
  {2008})}\BibitemShut {NoStop}%
\bibitem [{\citenamefont {Gambardella}\ \emph {et~al.}(2003)\citenamefont
  {Gambardella}, \citenamefont {Rusponi}, \citenamefont {Veronese},
  \citenamefont {Dhesi}, \citenamefont {Grazioli}, \citenamefont {Dallmeyer},
  \citenamefont {Cabria}, \citenamefont {Zeller}, \citenamefont {Dederichs},
  \citenamefont {Kern}, \citenamefont {Carbone},\ and\ \citenamefont
  {Brune}}]{gambardella2004giant}%
  \BibitemOpen
  \bibfield  {author} {\bibinfo {author} {\bibfnamefont {P.}~\bibnamefont
  {Gambardella}}, \bibinfo {author} {\bibfnamefont {S.}~\bibnamefont
  {Rusponi}}, \bibinfo {author} {\bibfnamefont {M.}~\bibnamefont {Veronese}},
  \bibinfo {author} {\bibfnamefont {S.}~\bibnamefont {Dhesi}}, \bibinfo
  {author} {\bibfnamefont {C.}~\bibnamefont {Grazioli}}, \bibinfo {author}
  {\bibfnamefont {A.}~\bibnamefont {Dallmeyer}}, \bibinfo {author}
  {\bibfnamefont {I.}~\bibnamefont {Cabria}}, \bibinfo {author} {\bibfnamefont
  {R.}~\bibnamefont {Zeller}}, \bibinfo {author} {\bibfnamefont
  {P.}~\bibnamefont {Dederichs}}, \bibinfo {author} {\bibfnamefont
  {K.}~\bibnamefont {Kern}}, \bibinfo {author} {\bibfnamefont {C.}~\bibnamefont
  {Carbone}}, \ and\ \bibinfo {author} {\bibfnamefont {H.}~\bibnamefont
  {Brune}},\ }\href@noop {} {\bibfield  {journal} {\bibinfo  {journal}
  {Science}\ }\textbf {\bibinfo {volume} {300}},\ \bibinfo {pages} {1130}
  (\bibinfo {year} {2003})}\BibitemShut {NoStop}%
\bibitem [{\citenamefont {Hirjibehedin}\ \emph {et~al.}(2007)\citenamefont
  {Hirjibehedin}, \citenamefont {Lin}, \citenamefont {Otte}, \citenamefont
  {Ternes}, \citenamefont {Lutz}, \citenamefont {Jones},\ and\ \citenamefont
  {Heinrich}}]{hirjibehedin2007anisotropy}%
  \BibitemOpen
  \bibfield  {author} {\bibinfo {author} {\bibfnamefont {C.~F.}\ \bibnamefont
  {Hirjibehedin}}, \bibinfo {author} {\bibfnamefont {C.-Y.}\ \bibnamefont
  {Lin}}, \bibinfo {author} {\bibfnamefont {A.~F.}\ \bibnamefont {Otte}},
  \bibinfo {author} {\bibfnamefont {M.}~\bibnamefont {Ternes}}, \bibinfo
  {author} {\bibfnamefont {C.~P.}\ \bibnamefont {Lutz}}, \bibinfo {author}
  {\bibfnamefont {B.~A.}\ \bibnamefont {Jones}}, \ and\ \bibinfo {author}
  {\bibfnamefont {A.~J.}\ \bibnamefont {Heinrich}},\ }\href@noop {} {\bibfield
  {journal} {\bibinfo  {journal} {Science}\ }\textbf {\bibinfo {volume}
  {317}},\ \bibinfo {pages} {1199} (\bibinfo {year} {2007})}\BibitemShut
  {NoStop}%
\bibitem [{\citenamefont {Khajetoorians}\ \emph {et~al.}(2013)\citenamefont
  {Khajetoorians}, \citenamefont {Schlenk}, \citenamefont {Schweflinghaus},
  \citenamefont {dos Santos~Dias}, \citenamefont {Steinbrecher}, \citenamefont
  {Bouhassoune}, \citenamefont {Lounis}, \citenamefont {Wiebe},\ and\
  \citenamefont {Wiesendanger}}]{khajetoorians2013spin}%
  \BibitemOpen
  \bibfield  {author} {\bibinfo {author} {\bibfnamefont {A.}~\bibnamefont
  {Khajetoorians}}, \bibinfo {author} {\bibfnamefont {T.}~\bibnamefont
  {Schlenk}}, \bibinfo {author} {\bibfnamefont {B.}~\bibnamefont
  {Schweflinghaus}}, \bibinfo {author} {\bibfnamefont {M.}~\bibnamefont {dos
  Santos~Dias}}, \bibinfo {author} {\bibfnamefont {M.}~\bibnamefont
  {Steinbrecher}}, \bibinfo {author} {\bibfnamefont {M.}~\bibnamefont
  {Bouhassoune}}, \bibinfo {author} {\bibfnamefont {S.}~\bibnamefont {Lounis}},
  \bibinfo {author} {\bibfnamefont {J.}~\bibnamefont {Wiebe}}, \ and\ \bibinfo
  {author} {\bibfnamefont {R.}~\bibnamefont {Wiesendanger}},\ }\href@noop {}
  {\bibfield  {journal} {\bibinfo  {journal} {Phys. Rev. Lett,}\ }\textbf
  {\bibinfo {volume} {111}},\ \bibinfo {pages} {157204} (\bibinfo {year}
  {2013})}\BibitemShut {NoStop}%
\bibitem [{\citenamefont {Bryant}\ \emph {et~al.}(2013)\citenamefont {Bryant},
  \citenamefont {Spinelli}, \citenamefont {Wagenaar}, \citenamefont {Gerrits},\
  and\ \citenamefont {Otte}}]{bryant2013local}%
  \BibitemOpen
  \bibfield  {author} {\bibinfo {author} {\bibfnamefont {B.}~\bibnamefont
  {Bryant}}, \bibinfo {author} {\bibfnamefont {A.}~\bibnamefont {Spinelli}},
  \bibinfo {author} {\bibfnamefont {J.}~\bibnamefont {Wagenaar}}, \bibinfo
  {author} {\bibfnamefont {M.}~\bibnamefont {Gerrits}}, \ and\ \bibinfo
  {author} {\bibfnamefont {A.}~\bibnamefont {Otte}},\ }\href@noop {} {\bibfield
   {journal} {\bibinfo  {journal} {Phys. Rev. Lett.}\ }\textbf {\bibinfo
  {volume} {111}},\ \bibinfo {pages} {127203} (\bibinfo {year}
  {2013})}\BibitemShut {NoStop}%
\bibitem [{\citenamefont {Heinrich}\ \emph {et~al.}(2013)\citenamefont
  {Heinrich}, \citenamefont {Braun}, \citenamefont {Pascual},\ and\
  \citenamefont {Franke}}]{heinrich2013protection}%
  \BibitemOpen
  \bibfield  {author} {\bibinfo {author} {\bibfnamefont {B.}~\bibnamefont
  {Heinrich}}, \bibinfo {author} {\bibfnamefont {L.}~\bibnamefont {Braun}},
  \bibinfo {author} {\bibfnamefont {J.}~\bibnamefont {Pascual}}, \ and\
  \bibinfo {author} {\bibfnamefont {K.}~\bibnamefont {Franke}},\ }\href@noop {}
  {\bibfield  {journal} {\bibinfo  {journal} {Nature Physics}\ }\textbf
  {\bibinfo {volume} {9}},\ \bibinfo {pages} {765} (\bibinfo {year}
  {2013})}\BibitemShut {NoStop}%
\bibitem [{\citenamefont {Loth}\ \emph {et~al.}(2010)\citenamefont {Loth},
  \citenamefont {Etzkorn}, \citenamefont {Lutz}, \citenamefont {Eigler},\ and\
  \citenamefont {Heinrich}}]{loth2010measurement}%
  \BibitemOpen
  \bibfield  {author} {\bibinfo {author} {\bibfnamefont {S.}~\bibnamefont
  {Loth}}, \bibinfo {author} {\bibfnamefont {M.}~\bibnamefont {Etzkorn}},
  \bibinfo {author} {\bibfnamefont {C.~P.}\ \bibnamefont {Lutz}}, \bibinfo
  {author} {\bibfnamefont {D.}~\bibnamefont {Eigler}}, \ and\ \bibinfo {author}
  {\bibfnamefont {A.~J.}\ \bibnamefont {Heinrich}},\ }\href@noop {} {\bibfield
  {journal} {\bibinfo  {journal} {Science}\ }\textbf {\bibinfo {volume}
  {329}},\ \bibinfo {pages} {1628} (\bibinfo {year} {2010})}\BibitemShut
  {NoStop}%
\bibitem [{\citenamefont {MacKay}(2003)}]{mackay2003information}%
  \BibitemOpen
  \bibfield  {author} {\bibinfo {author} {\bibfnamefont {D.~J.}\ \bibnamefont
  {MacKay}},\ }\href@noop {} {\emph {\bibinfo {title} {Information theory,
  inference and learning algorithms}}}\ (\bibinfo  {publisher} {Cambridge
  university press},\ \bibinfo {year} {2003})\BibitemShut {NoStop}%
\bibitem [{\citenamefont {Kim}\ and\ \citenamefont
  {Huse}(2013)}]{kim2013ballistic}%
  \BibitemOpen
  \bibfield  {author} {\bibinfo {author} {\bibfnamefont {H.}~\bibnamefont
  {Kim}}\ and\ \bibinfo {author} {\bibfnamefont {D.~A.}\ \bibnamefont {Huse}},\
  }\href@noop {} {\bibfield  {journal} {\bibinfo  {journal} {Phys. Rev. Lett.}\
  }\textbf {\bibinfo {volume} {111}},\ \bibinfo {pages} {127205} (\bibinfo
  {year} {2013})}\BibitemShut {NoStop}%
\bibitem [{\citenamefont {Kitaev}(2001)}]{kitaev2001unpaired}%
  \BibitemOpen
  \bibfield  {author} {\bibinfo {author} {\bibfnamefont {A.~Y.}\ \bibnamefont
  {Kitaev}},\ }\href@noop {} {\bibfield  {journal} {\bibinfo  {journal}
  {Physics-Uspekhi}\ }\textbf {\bibinfo {volume} {44}},\ \bibinfo {pages} {131}
  (\bibinfo {year} {2001})}\BibitemShut {NoStop}%
\bibitem [{\citenamefont {Kitaev}\ and\ \citenamefont
  {Laumann}(2009)}]{kitaev2009topological}%
  \BibitemOpen
  \bibfield  {author} {\bibinfo {author} {\bibfnamefont {A.}~\bibnamefont
  {Kitaev}}\ and\ \bibinfo {author} {\bibfnamefont {C.}~\bibnamefont
  {Laumann}},\ }\href@noop {} {\bibfield  {journal} {\bibinfo  {journal}
  {arXiv:0904.2771}\ } (\bibinfo {year} {2009})}\BibitemShut {NoStop}%
\bibitem [{\citenamefont {Blaizot}\ and\ \citenamefont
  {Ripka}(1986)}]{blaizot1986quantum}%
  \BibitemOpen
  \bibfield  {author} {\bibinfo {author} {\bibfnamefont {J.}~\bibnamefont
  {Blaizot}}\ and\ \bibinfo {author} {\bibfnamefont {G.}~\bibnamefont
  {Ripka}},\ }\href@noop {} {\emph {\bibinfo {title} {Quantum Theory of Finite
  Systems}}}\ (\bibinfo  {publisher} {Cambridge, MA},\ \bibinfo {year}
  {1986})\BibitemShut {NoStop}%
\end{thebibliography}
\end{document}